\documentclass[prd,notitlepage,longbibliography,nofootinbib,superscriptaddress,onecolumn,preprintnumbers]{revtex4-2}
\usepackage[utf8]{inputenc}

\usepackage{bm}
\usepackage{comment} 
\usepackage[colorlinks=true,urlcolor=blue,anchorcolor=black,citecolor=blue,linkcolor=red,filecolor=black,menucolor=black,pagecolor=black,linktocpage=true,pdfproducer=medialab,pdfa=true]{hyperref}
\usepackage{graphicx}
\usepackage{amsmath,latexsym,amssymb,mathrsfs,ascmac,mathtools}
\usepackage{multirow}
\usepackage{braket}
\usepackage{placeins}
\usepackage{float}
\usepackage{subcaption}
\usepackage{makecell}
\begin{document}

\title{Gravitational Lensing of Hayward Black Holes with EFT-Corrected Photon Propagation}

\author{Takamasa Kanai}
\email{kanai@kochi-ct.ac.jp}

\affiliation{Department of Social Design Engineering,
National Institute of Technology (KOSEN), Kochi College,
200-1 Monobe Otsu, Nankoku, Kochi, 783-8508, Japan}

\begin{abstract}
We investigate whether a Hayward regular black hole can be observationally distinguished from a Schwarzschild black hole through strong gravitational lensing when effective field theory (EFT) corrections to photon propagation are taken into account. We derive the modified photon propagation law induced by non-minimal couplings between the electromagnetic field and spacetime curvature, and analyze the resulting photon trajectories in the Hayward spacetime. Using the strong deflection limit, we derive the corrections to the photon sphere and the logarithmically divergent part of the deflection angle. Although the EFT corrections are parametrically small, their effects can be enhanced near the critical propagation region, where the deflection angle exhibits a logarithmic divergence. We evaluate the strong-deflection observables for representative values of the Hayward parameter and compare them with the Schwarzschild case. We find that EFT corrections to photon propagation can leave characteristic imprints on strong-lensing observables. Furthermore, the contribution of the EFT corrections becomes more pronounced as the Hayward parameter $g^3$ approaches its critical value, indicating that curvature-dependent corrections to photon propagation can become particularly relevant in the near-critical regime. These results suggest that strong gravitational lensing may provide a means of distinguishing Hayward regular black holes from Schwarzschild black holes.
\end{abstract}
\maketitle

\section{Introduction}

Gravitational phenomena in the strong-field regime provide a unique window into the nature of gravity and the structure of compact objects beyond the weak-field approximation. In particular, the propagation of light near unstable photon orbits gives rise to characteristic observables such as black hole shadows \cite{Falcke:1999pj,EventHorizonTelescope:2019dse,EventHorizonTelescope:2022wkp,Vagnozzi:2022moj} and gravitational lensing \cite{Virbhadra:1999nm,Bozza:2001xd,Bozza:2002zj,Gibbons:2008rj}. These observables provide promising probes for testing the nature of compact objects and searching for deviations from the Schwarzschild black hole predicted by general relativity.

Among these probes, strong gravitational lensing is particularly sensitive to the geometry in the vicinity of an unstable photon orbit. In the strong deflection limit, the deflection angle develops a logarithmic divergence as the closest approach radius approaches the critical radius associated with the unstable photon orbit. This behavior is captured by the formalism developed by Bozza \cite{Bozza:2002zj,Bozza:2002af,Bozza:2003cp}, in which the deflection angle is characterized by a logarithmically divergent term and a finite regular contribution. Since the coefficients of this expansion are determined by the local properties of the effective geometry near the critical orbit, strong gravitational lensing can be particularly sensitive to small deviations from the underlying spacetime geometry.

This sensitivity makes strong gravitational lensing a useful probe of alternative compact-object geometries, including regular black holes \cite{1968qtr..conf...87B,Ayon-Beato:1998hmi,Ayon-Beato:1999kuh,Hayward:2005gi,Ashtekar:2005qt,Modesto:2005zm}, whose strong-lensing and shadow signatures have been investigated in various contexts \cite{Ghaffarnejad:2014zva,Zhang:2018yzr,Gera:2024qob,Liu:2026yyj,Yuan:2026qxe}. Unlike the Schwarzschild black hole, regular black holes are constructed to avoid the curvature singularity at the center while retaining a black-hole-like exterior geometry. They therefore provide phenomenological models in which the strong-field structure can differ from that of the Schwarzschild black hole while remaining regular at the center. Among various regular black hole models, the Hayward black hole \cite{Hayward:2005gi} provides a simple and widely studied example, in which the Schwarzschild geometry is recovered in the appropriate limit. The Hayward parameter modifies the geometry in the strong-field region and consequently changes the properties of unstable photon orbits and the associated gravitational-lensing observables. This naturally raises the question of whether a Hayward regular black hole can be observationally distinguished from the Schwarzschild black hole through strong gravitational lensing.

In addition to the modifications of the background geometry, the propagation of photons can itself receive corrections within the effective field theory (EFT) \cite{Weinberg:1978kz,Donoghue:1993eb,Donoghue:1994dn,Burgess:2003jk}. EFT provides a systematic framework for describing corrections to general relativity at low energies, including higher-curvature interactions and non-minimal couplings between the electromagnetic field and spacetime curvature. Such curvature-photon interactions modify the photon dispersion relation and, consequently, the propagation of light in curved spacetime. In particular, the modified propagation law can depend on the photon polarization and give rise to gravitational birefringence \cite{Scharnhorst:1990sr,Barton:1989dq,Barton:1992pq,Latorre:1994cv,Dittrich:1998fy,DeLorenci:2000yh,Drummond:1979pp,Daniels:1993yi,Shore:1995fz,Daniels:1995yw,Shore:2007um,Cho:1997vg,Izumi:2014loa,Reall:2014pwa,Allahyari:2019jqz,Cao:2021sty,Reall:2021voz,Davies:2021frz,Fu:2025oxr,Kanai:2026ohg}. Consequently, when strong gravitational lensing is used to distinguish different compact-object geometries, the relevant photon trajectories and strong-deflection observables are determined not only by the background spacetime but also by the effective propagation law of photons. This motivates the study of strong gravitational lensing in Hayward spacetime while consistently taking into account both the modified background geometry and the EFT corrections to photon propagation.

A particularly relevant feature of the Hayward spacetime is that, unlike the Schwarzschild vacuum, it is generally non-Ricci-flat. Consequently, curvature-photon interactions involving the Ricci tensor can contribute to photon propagation in the Hayward background while vanishing in the Schwarzschild spacetime. The Weyl curvature also contributes to the modified photon propagation through the corresponding curvature-photon interaction. Thus, EFT corrections provide an additional way to probe the curvature structure of the background spacetime beyond the modification of the metric itself. This raises the possibility that the combination of the Hayward geometry and curvature-dependent photon propagation may leave characteristic signatures in strong-lensing observables.

Even when the corrections to the background geometry and photon propagation are perturbatively small, their effects on strong-lensing observables need not remain negligible. In the strong deflection regime, the logarithmic divergence of the deflection angle reflects the prolonged propagation of photons near the critical orbit and can enhance the sensitivity to small corrections. Therefore, strong gravitational lensing provides a potentially useful probe for identifying subtle differences between regular and Schwarzschild black holes and for investigating the effects of EFT corrections to photon propagation.

In this work, we investigate strong gravitational lensing by a Hayward regular black hole while taking into account EFT corrections to photon propagation. We first analyze photon propagation and strong gravitational lensing in the uncorrected Hayward spacetime, deriving the photon-sphere radius, critical impact parameter, deflection angle, and corresponding strong-deflection observables. We then incorporate the curvature-dependent EFT corrections and derive the modified photon propagation law and the corresponding effective optical metrics for the physical polarization modes. The resulting photon trajectories depend on the polarization, and the critical photon orbits and strong-deflection coefficients are consequently modified by the EFT corrections.

We evaluate the strong-deflection observables for representative values of the Hayward parameter and compare them with the corresponding Schwarzschild results. In particular, we examine how the EFT corrections modify the deflection angle and photon propagation time, and whether these modifications leave characteristic signatures that can distinguish the Hayward regular black hole from the Schwarzschild case. By taking into account both the difference in the background geometry and the curvature-dependent modification of photon propagation, we assess the potential of strong gravitational lensing as a probe for distinguishing regular black holes from Schwarzschild black holes.

In Sec.~\ref{sec:HB}, we introduce the effective action for the Hayward black hole and derive the modified photon propagation law induced by the curvature-photon couplings, together with the corresponding effective metric. In Sec.~\ref{sec:HBlens}, we review gravitational lensing in the Hayward black hole spacetime in the absence of EFT corrections, deriving the photon-sphere radius, critical impact parameter, and the strong-deflection expansion of the deflection angle, which serve as the reference results for the subsequent analysis. In Sec.~\ref{sec:strong lens}, we investigate gravitational lensing in the strong-deflection limit and derive the effects of the EFT corrections on the photon sphere and the deflection angle. We further discuss the characteristic features of the lensing observables in the Hayward black hole spacetime and highlight their differences from those expected in the Schwarzschild black hole spacetime. In Sec.~\ref{sec:time delay}, we study the photon time delay in the strong-deflection regime and examine its sensitivity to the EFT corrections. Finally, in Sec.~\ref{sec:conclusion}, we summarize our results and discuss their implications and future prospects.

In this paper, we set the Newton constant and the speed of light equal to unity.

\section{Photon Propagation and Effective Geometry in EFT-Corrected Hayward black hole}
\label{sec:HB}

Gravitational lensing by compact objects provides a powerful probe of strong-field gravity and the underlying spacetime geometry~\cite{Einstein:1936llh,Virbhadra:1999nm,Bozza:2001xd,Bozza:2002zj,Gibbons:2008rj}. Beyond the geometry of the background spacetime, the propagation of light can also be modified by non-minimal couplings between the electromagnetic field and spacetime curvature. Such curvature couplings arise naturally from quantum corrections to electrodynamics in curved spacetime~\cite{Drummond:1979pp} and can lead to polarization-dependent photon propagation, giving rise to a gravitational analogue of birefringence~\cite{Chen:2015cpa,Lu:2016gsf}. Among the possible curvature couplings, the interaction between the electromagnetic field and the Weyl tensor has been extensively studied, including its effects on photon propagation and gravitational lensing in black-hole spacetimes~\cite{Drummond:1979pp,Chen:2015cpa,Lu:2016gsf,Chen:2016hil}.

In this work, we investigate a broader class of curvature-photon couplings, including both Weyl and Ricci tensor interactions, in the context of a Hayward regular black hole. We study how these EFT corrections modify photon propagation and whether they can lead to observable differences between the Hayward and Schwarzschild black holes through strong gravitational lensing. In contrast to the minimally coupled case, the curvature-photon interactions modify the photon propagation law and can lead to polarization-dependent photon trajectories. We derive the corresponding effective optical metrics from the photon dispersion relation in the eikonal approximation and use them to analyze the strong-deflection gravitational lensing in the Hayward spacetime. In particular, we determine the corrections to the photon-sphere radius and the logarithmically divergent part of the deflection angle, and investigate how these corrections depend on the Hayward parameter and the EFT couplings. This framework allows us to assess whether the combined effects of the nontrivial regular black-hole geometry and curvature-dependent photon propagation can leave characteristic signatures that distinguish a Hayward regular black hole from the Schwarzschild black hole.

\subsection{Equation of motion for a higher curvature-coupled photon}

The starting point of our analysis is a gravitational theory described by an $f(R)$ term and a scalar field, supplemented by effective field theory (EFT) corrections that modify photon propagation. We consider the effective action

\begin{align}
S=\int d^{4}x\sqrt{-g}\Biggl[
\frac{1}{\kappa^2}
\left(
f(R)+\psi(\chi)\nabla_{\mu}\chi\nabla^{\mu}\chi
\right)
-\frac{1}{4}F_{\mu\nu}F^{\mu\nu}
+\alpha R F_{\mu\nu}F^{\mu\nu}
+\beta R_{\mu\nu}F^{\mu\rho}F^{\nu}{}{\rho}
+\gamma C^{\mu\nu\rho\sigma}F{\mu\nu}F_{\rho\sigma}
\Biggr],
\label{eq:action}
\end{align}
where $F_{\mu\nu}$ denotes the electromagnetic field strength, $R_{\mu\nu}$ and $R$ are the Ricci tensor and Ricci scalar, respectively, $C_{\mu\nu\rho\sigma}$ is the Weyl tensor, $f$ is a function of the Ricci scalar $R$, and $\alpha$, $\beta$, and $\gamma$ are EFT coupling constants with dimensions of length squared.

We regard the higher-derivative operators as terms in a derivative expansion of the underlying effective theory. In this work, we restrict the analysis to the leading nontrivial order in this expansion and retain only operators containing four derivatives in total. Higher-derivative operators are therefore consistently neglected, as their effects are suppressed by additional powers of the EFT scale. The corresponding corrections are treated perturbatively, and only terms at first order in the EFT coupling constants are retained throughout the analysis.

The first term in the action describes the gravitational background through an $f(R)$ theory coupled to the scalar field $\chi$. The electromagnetic field is minimally coupled to this background through the standard Maxwell term. The remaining terms describe non-minimal interactions between the electromagnetic field and spacetime curvature. In particular, the $\beta$ and $\gamma$ terms couple the electromagnetic field to the Ricci and Weyl tensors, respectively, and modify the photon propagation law in the curved spacetime. The $\alpha$ term involving the Ricci scalar is also included for completeness.

We do not include direct couplings between the electromagnetic field and the scalar field, such as operators involving $\chi$ and $F_{\mu\nu}$. Instead, we describe the effect of the scalar sector through the background geometry determined by the coupled $f(R)$ gravity and scalar-field equations. In the exterior region of the black hole, where the curvature becomes weaker away from the horizon, higher-order curvature contributions are suppressed. We therefore focus on the leading curvature-dependent contribution to photon propagation. Using the background equations of motion, the contribution of the scalar field to photon propagation can, at the leading order of the curvature expansion, be approximately related to the Ricci curvature of the background spacetime. Accordingly, we represent this leading contribution by the curvature-photon interaction $R_{\mu\nu}F^{\mu\rho}F^{\nu}_{\ \rho}$, rather than introducing an independent direct scalar-photon coupling. This approximation captures the leading effect of the modified geometry determined by the coupled $f(R)$ gravity and scalar field, while consistently neglecting higher-order curvature contributions and additional model-dependent scalar-photon interactions.

Since the background electromagnetic field is assumed to vanish, $F_{\mu\nu}=0$, the curvature-photon interaction terms do not modify the background geometry at first order in the EFT couplings. Nevertheless, they contribute directly to the photon equations of motion and modify the photon dispersion relation at leading order in the eikonal approximation. These curvature-dependent interactions are therefore the relevant EFT operators for studying the propagation of photons and their gravitational lensing in the background spacetime considered below.

We take the Hayward regular black hole as the background spacetime. Unlike the Schwarzschild spacetime, the Hayward geometry is generally non-Ricci-flat, and its curvature structure is characterized by the Hayward parameter. Consequently, both Ricci- and Weyl-curvature contributions can affect photon propagation through the EFT interactions. In particular, the $\beta$ term probes the Ricci curvature of the Hayward background, while the $\gamma$ term couples the electromagnetic field to the Weyl curvature and therefore captures the effect of the gravitational curvature on photon propagation. This provides a natural framework for investigating how curvature-dependent photon propagation can distinguish a Hayward regular black hole from the Schwarzschild black hole.

Varying the action~\eqref{eq:action} with respect to the electromagnetic field gives the modified Maxwell equation,
\begin{align}
\nabla_{\mu}\left(F^{\mu\nu}-4\alpha R F^{\mu\nu}-2\beta R_{\mu\rho}F^{\rho\nu}-2\beta R^{\nu\rho}F^{\mu}{}{\rho}-4\gamma C^{\mu\nu\rho\sigma}F{\rho\sigma}\right)=0 .
\label{eq:maxwell-mod}
\end{align}
To study photon propagation in the geometric-optics regime, we employ the eikonal approximation and write the electromagnetic potential in terms of a rapidly varying phase and a slowly varying amplitude. Denoting the wave vector by $k_\mu=\nabla_\mu\theta$ and the polarization vector by $a^\mu$, we impose the transversality condition $k_\mu a^\mu=0$. Substituting the eikonal ansatz into the modified Maxwell equation and retaining the leading-order terms in the eikonal expansion, we obtain the following equation governing photon propagation:
\begin{align}
(1-4\alpha R)k_{\mu}k^{\mu}a^{\nu}-2\beta R_{\mu\rho}k^\mu k^\rho a^\nu+2\beta R_{\mu\rho}k^\mu a^\rho k^\nu-2\beta R^{\rho\nu}a_\rho k_{\mu}k^{\mu}+8\gamma C^{\mu\nu\rho\sigma}k_{\sigma}k_{\mu}a_{\rho}=0 .
\label{eq:eom-photon}
\end{align}
Since we treat the EFT corrections perturbatively, the zeroth-order propagation is governed by the GR null condition $k_\mu k^\mu=0$. Consequently, the terms proportional to $k_\mu k^\mu$, namely the term proportional to $\alpha R$ and the fourth term in Eq.~\eqref{eq:eom-photon}, do not contribute to the dispersion relation at first order in the EFT couplings. We therefore omit these terms in the following analysis.

\subsection{General static and spherically symmetric background and the Weyl tensor in an orthonormal frame}

The metric of a general static and spherically symmetric spacetime can be written as
\begin{align}
ds^{2}=-h(r)dt^{2}+\frac{dr^2}{\ell(r)}
+r^2\left(d\theta^{2}+\sin^2\theta,d\phi^2\right),
\label{eq:general-metric}
\end{align}
where $h(r)$ and $\ell(r)$ are arbitrary functions of the radial coordinate. This metric provides a general framework for studying photon propagation in static and spherically symmetric spacetimes, including the Hayward regular black hole considered below.

Following the analysis of~\cite{Daniels:1993yi,Shore:1995fz,Daniels:1995yw,Chen:2016hil,Kanai:2026ohg}, we introduce an orthonormal vierbein $e^{a}_{\ \mu}$ defined by
$g_{\mu\nu}=\eta_{ab}e^{a}_{\ \mu}e^{b}_{\ \nu}$, together with the antisymmetric combination
\begin{align}
U^{ab}_{\ \ \mu\nu}
=e^{a}_{\ \mu}e^{b}_{\ \nu}
-e^{a}_{\ \nu}e^{b}_{\ \mu}.
\end{align}
In this orthonormal frame, the Weyl tensor can be expressed in the compact bilinear form
\begin{align}
C_{\mu\nu\rho\sigma}
=\mathcal{A}\Big(
U^{01}_{\mu\nu}U^{01}_{\rho\sigma}
-U^{23}_{\mu\nu}U^{23}_{\rho\sigma}\Big)
+\mathcal{B}\Big(
U^{02}_{\mu\nu}U^{02}_{\rho\sigma}
+U^{03}_{\mu\nu}U^{03}_{\rho\sigma}
-U^{12}_{\mu\nu}U^{12}_{\rho\sigma}
-U^{13}_{\mu\nu}U^{13}_{\rho\sigma}
\Big),
\label{eq:weyl-general}
\end{align}
where $\mathcal{A}$ and $\mathcal{B}$ are scalar functions that depend only on $r$ and the metric functions $h(r)$ and $\ell(r)$ and their derivatives. Their explicit forms are
\begin{align}
\mathcal{A}
&=
-\frac{
r^2h'^2\ell
+h(4-4\ell-r\ell')
-rh\left(rh'\ell'+\ell(h'+2rh'')\right)
}{6r^2h^2},
\label{A}\\
\mathcal{B}
&=
-\frac{
r^2h'^2\ell
+2h^2(2-2\ell-5r\ell')
-rh\left(rh'\ell+2\ell(5h'+rh'')\right)
}{24r^2h^2}.
\label{B}
\end{align}

These expressions are valid for a general static and spherically symmetric spacetime and will be used below to derive the modified photon propagation law in the presence of the Weyl-photon coupling.

\subsection{Polarization decomposition and the light-cone condition}

To organize the photon equation of motion~\eqref{eq:eom-photon} in a general static and spherically symmetric background, we introduce the momentum-projected combinations~\cite{Daniels:1993yi,Shore:1995fz,Daniels:1995yw}
\begin{equation}
l_{\nu}=k^{\mu}U^{01}_{\ \ \mu\nu},\qquad
n_{\nu}=k^{\mu}U^{02}_{\ \ \mu\nu},\qquad
m_{\nu}=k^{\mu}U^{23}_{\ \ \mu\nu},
\label{eq:lnr}
\end{equation}
together with the related combinations
\begin{equation}
p_{\nu}=k^{\mu}U^{12}_{\ \ \mu\nu},\qquad
q_{\nu}=k^{\mu}U^{03}_{\ \ \mu\nu},\qquad
r_{\nu}=k^{\mu}U^{13}_{\ \ \mu\nu},
\end{equation}
all of which are orthogonal to $k^\nu$. Contracting Eq.~\eqref{eq:eom-photon} with the appropriate independent polarization vectors reduces the photon equation of motion to a homogeneous linear system for the three independent polarization amplitudes $a\cdot l$, $a\cdot n$, and $a\cdot m$,
\begin{equation}
\begin{pmatrix}
K_{11} & 0 & 0\\
K_{21} & K_{22} & K_{23}\\
0 & 0 & K_{33}
\end{pmatrix}
\begin{pmatrix}
a\cdot l\\
a\cdot n\\
a\cdot m
\end{pmatrix}
=0.
\label{eq:Kmatrix}
\end{equation}
The coefficients $K_{ij}$ depend on the metric functions, the curvature tensors, and the components of the wave vector. At the order relevant to the present perturbative analysis, they are given by
\begin{align}
K_{11}={}&
(1+8\gamma\mathcal{A})
(g_{00}k^0k^0+g_{11}k^1k^1)
+(1+8\gamma\mathcal{B})
(g_{22}k^2k^2+g_{33}k^3k^3)
-2\beta R_{\mu\nu}k^\mu k^\nu,
\\
K_{21}={}&
8\gamma(\mathcal{A}-\mathcal{B})
\sqrt{g_{11}g_{22}}\,k^1k^2,
\\
K_{22}={}&
(1+8\gamma\mathcal{B})
(g_{00}k^0k^0+g_{11}k^1k^1
+g_{22}k^2k^2+g_{33}k^3k^3)
-2\beta R_{\mu\nu}k^\mu k^\nu,
\\
K_{23}={}&
-8\gamma(\mathcal{A}-\mathcal{B})
\sqrt{-g_{00}g_{33}}\,k^0k^3,
\\
K_{33}={}&
(1+8\gamma\mathcal{B})
(g_{00}k^0k^0+g_{11}k^1k^1)
+(1+8\gamma\mathcal{A})
(g_{22}k^2k^2+g_{33}k^3k^3)
-2\beta R_{\mu\nu}k^\mu k^\nu.
\end{align}

For a nontrivial solution, the determinant of the coefficient matrix must vanish. At the order considered here, the determinant factorizes into three propagation conditions,
\begin{align}
\det K=\tilde K_{11}\tilde K_{22}\tilde K_{33}=0.
\label{eq:Kdet-eq}
\end{align}
The first condition, $\tilde K_{11}=0$, gives
\begin{align}
&\left[(1+8\gamma\mathcal{A})g_{00}-2\beta R_{00}\right]k^0k^0
+\left[(1+8\gamma\mathcal{A})g_{11}-2\beta R_{11}\right]k^1k^1
\notag\\
&\quad
+\left[(1+8\gamma\mathcal{B})g_{22}-2\beta R_{22}\right]k^2k^2
+\left[(1+8\gamma\mathcal{B})g_{33}-2\beta R_{33}\right]k^3k^3
=0.
\label{eq:PPL}
\end{align}
This mode corresponds to a polarization vector lying within the orbital plane. Following Refs.~\cite{Chen:2015cpa,Lu:2016gsf}, we refer to this polarization as the \emph{PPL} mode.

The second condition, $\tilde K_{22}=0$, reduces to the standard null condition
\begin{equation}
g_{\mu\nu}k^\mu k^\nu=0.
\end{equation}
It corresponds to the non-propagating polarization branch in the curvature-coupled system and does not represent an independent physical polarization mode. We therefore discard this branch, following the treatment in Refs.~\cite{Drummond:1979pp,Chen:2015cpa,Lu:2016gsf}.

The third condition, $\tilde K_{33}=0$, is given by
\begin{align}
&\left[(1+8\gamma\mathcal{B})g_{00}-2\beta R_{00}\right]k^0k^0
+\left[(1+8\gamma\mathcal{B})g_{11}-2\beta R_{11}\right]k^1k^1
\notag\\
&\quad
+\left[(1+8\gamma\mathcal{A})g_{22}-2\beta R_{22}\right]k^2k^2
+\left[(1+8\gamma\mathcal{A})g_{33}-2\beta R_{33}\right]k^3k^3
=0.
\label{eq:PPM}
\end{align}
This mode corresponds to a polarization vector orthogonal to the orbital plane and is denoted by \emph{PPM} following Ref.~\cite{Chen:2016hil}.

Equations~\eqref{eq:PPL} and~\eqref{eq:PPM} therefore describe the two physical polarization-dependent propagation modes of curvature-coupled photons in the equatorial plane of a general static and spherically symmetric spacetime. They define two distinct effective light cones and, equivalently, two effective optical metrics. Thus, photons with different polarization states can follow different trajectories, giving rise to a gravitational analogue of birefringence. In the limit $\beta,\gamma\rightarrow0$, both propagation conditions reduce to the standard null condition of the background spacetime, and the distinction between the two polarization modes disappears.

We now specialize to the modified Hayward regular black hole \cite{Nashed:2024lem} as the background spacetime. The Hayward geometry provides a simple model of a nonsingular black hole and is characterized by a length scale associated with the regularization of the central  region. We consider the static and spherically symmetric metric
\begin{align}
\label{hayward metric}
ds^2=-h(r)dt^2+\frac{dr^2}{\ell(r)}
+r^2(d\theta^2+\sin^2\theta\,d\phi^2),
\end{align}
where, for the modified Hayward black hole considered here,
\begin{align}
\label{f_metric}
h(r)&=1-\frac{2Mr^2}{r^3+g^3},\\
\label{g_metric}
\ell(r)&=h(r)e^{\alpha r^3},
\end{align}
where $M$ denotes the mass parameter and $g$ is a dimensional parameter with dimensions of length. The resulting spacetime is asymptotically Schwarzschild, while the geometry is regularized in the strong-field region by the Hayward parameter $g$.
This background will be used as the reference geometry for studying the effects of EFT corrections to photon propagation.

The curvature-photon interactions considered above modify the photon light-cone conditions in this background. For the subsequent analysis,
it is convenient to express the two physical propagation conditions, corresponding to the PPL and PPM polarization modes, in terms of effective
optical metrics. To first order in the EFT couplings, these effective metrics can be written in the unified form
\begin{align} 
\label{metric_1} 
ds^2=&-h(r)\left(1+\beta\mathcal{C}\right)^{-1}dt^2+\frac{1}{\ell(r)}\left(1+\beta\mathcal{D}\right)^{-1}dr^2+r^2\left(1+\beta\mathcal{E}\right)^{-1}\left(\dfrac{1+8\gamma\mathcal{A}}{1+8\gamma\mathcal{B}}\right)^s(d\theta^2+\sin^2\theta d\phi^2),\\ 
\mathcal{A}=&\frac{1}{6r^{2}\left(g^{3}+r^{3}\right)^{3}} \Bigg[ g^{9}\left(-4+e^{\alpha r^{3}} \left(4+3\alpha r^{3}\right)\right) +3g^{6}r^{2} \left( -4r+ e^{\alpha r^{3}}(-2M+r) \left(4+3\alpha r^{3}\right) \right)\notag\\ 
&+r^{8}\left(-4r+e^{\alpha r^{3}}\left(-12M+4r+3\alpha r^{4}\right)\right)+3g^{3}r^{5}\left(-4r+e^{\alpha r^{3}}\left(12M+4r-6\alpha Mr^{3}+3\alpha r^{4}\right)\right)\Bigg],\\ 
\mathcal{B}=&\frac{1}{12r^{2}\left(g^{3}+r^{3}\right)^{3}}\Biggl[3g^{3}r^{6}\left(-2+e^{\alpha r^{3}}\left(2-3\alpha\left(7M-5r\right)r^{2}\right)\right)+g^{9} \left(-2+e^{\alpha r^{3}}\left(2+15\alpha r^{3}\right)\right)\nonumber\\ 
&+r^{8}\left(-2r+e^{\alpha r^{3}}\left(12M+2r-27\alpha M r^{3}+15\alpha r^{4}\right)\right)+3g^{6}r^{2}\left(-2r+e^{\alpha r^{3}}\left(-4M\left(4+3\alpha r^{3}\right)+r\left(2+15\alpha r^{3}\right)\right)\right)\Biggr],\\ 
\mathcal{C}=&\dfrac{3Me^{\alpha r^3}(\alpha r^9-2g^6(2+\alpha r^3)+g^3(8r^3-\alpha r^6))}{(r^3+g^3)^3},\\ 
\mathcal{D}=&\dfrac{3e^{3\alpha r^3}(2\alpha g^9 r+\alpha r^9(2r-3M)+g^3(6 \alpha r^7-9\alpha Mr^6+8Mr^3)+g^6(6\alpha r^4-2M(3\alpha r^3+2))}{(r^3+g^3)^3},\\ 
\mathcal{E}=&-\dfrac{g^6(2-e^{-\alpha r^3}(2+3\alpha r^3))-2g^3 r^2(-2r+e^{\alpha r^3}(3\alpha r^4-3\alpha M r^3+2r-6M))+r^6(2-e^{\alpha r^3}(2+3\alpha r^2(r-2M)))}{r^2(r^3+g^3)^2}. 
\end{align}
Here, $h(r)$ and $\ell(r)$ are given in Eqs.~\eqref{f_metric} and \eqref{g_metric}, respectively, while $s=+1$ and $s=-1$ correspond to the PPL and PPM polarization modes, respectively. The functions $\mathcal{A}$ and $\mathcal{B}$ encode the Weyl-curvature contribution, while $\mathcal{C}$, $\mathcal{D}$, and $\mathcal{E}$ arise from the Ricci-curvature coupling. Thus, the polarization-mode dependence of the effective metric originates from the coupling between the photon and the Weyl curvature.

The effective metric reduces to the underlying Hayward background in the limit $\beta,\gamma\rightarrow0$. We treat the EFT corrections perturbatively and retain only terms at first order in the EFT coupling constants. The effective optical metrics in Eq.~\eqref{metric_1} therefore provide the basis for analyzing photon trajectories and strong gravitational lensing in the EFT-corrected Hayward regular black hole spacetime.

It is worth noting that the modified Hayward geometry with $\alpha\neq0$ exhibits a problematic asymptotic behavior. For $\alpha>0$, the curvature invariants grow without bound as $r\rightarrow\infty$, indicating that the spacetime does not possess a regular weak-curvature asymptotic region. For $\alpha<0$, although the curvature itself does not exhibit the same divergence, the lensing observables become divergent in the asymptotic region as $r\rightarrow\infty$. We omit the detailed analysis of these divergences here, as they are not directly relevant to the strong-field lensing region of interest. These features suggest that the modified Hayward geometry with $\alpha\neq0$ should not be regarded as a complete infrared description of the spacetime, but rather as a geometry incorporating modifications that may be relevant to the ultraviolet regime.

In the present work, we therefore restrict our attention to the case $\alpha=0$, for which the metric reduces to the standard Hayward regular black hole. This choice provides a well-behaved asymptotic region while retaining the characteristic nonsingular strong-field structure of the Hayward spacetime. Our subsequent analysis of the EFT corrections to photon propagation and gravitational lensing is thus performed on the standard Hayward background, allowing us to focus on the effects of the curvature-photon interactions without introducing additional asymptotic divergences associated with the $\alpha$-dependent modification.

\section{Gravitational Lensing in the Hayward Black Hole Spacetime}
\label{sec:HBlens}

In this section, we review the gravitational lensing of photons in the Hayward regular black hole spacetime without EFT corrections. The purpose of this section is to establish the standard strong-deflection behavior of the Hayward black hole and to provide a reference result for the analysis with EFT corrections presented in the subsequent section. We first determine the photon sphere and the corresponding critical impact parameter, and then derive the deflection angle in the strong-deflection limit. We also obtain the associated lensing observables and compare them with those of the Schwarzschild black hole. These results provide the baseline predictions for the Hayward regular black hole against which the effects of the curvature-photon interactions can be assessed.

We consider the propagation of minimally coupled photons in the Hayward background and analyze their trajectories in the equatorial plane. Following the standard strong-deflection formalism~\cite{Bozza:2002zj,Bozza:2002af,Bozza:2003cp}, we derive the logarithmically divergent behavior of the deflection angle as the closest approach radius approaches the photon-sphere radius. The deflection angle is expressed in terms of the logarithmically divergent contribution and a finite regular part, whose coefficients are determined by the properties of the Hayward geometry near the photon sphere. We then derive the corresponding strong-lensing observables and investigate their dependence on the Hayward parameter.

Finally, we compare the strong-deflection coefficients and the resulting lensing observables of the Hayward regular black hole with those of the Schwarzschild black hole. This comparison provides a reference for assessing whether the intrinsic differences between the two spacetimes can be reflected in gravitational-lensing observables. In the next section, we extend this analysis by incorporating the EFT-induced modifications to photon propagation and investigate how the resulting corrections alter the strong-deflection observables.

\subsection{Gravitational Lensing in the Hayward Black Hole}

We briefly review the gravitational lensing of photons in the Hayward black hole spacetime following Ref.~\cite{Chiba:2017nml}. For $\alpha=0$, the modified Hayward metric given in Eq.~(\ref{hayward metric}) reduces to the standard Hayward metric,
\begin{align}
    ds^2=-F(r)dt^2+\dfrac{dr^2}{F(r)}+r^2d\Omega^2,
\end{align}
where
\begin{align}
    F(r)=1-\frac{2Mr^2}{r^3+g^3}.
\end{align}
The metric approaches the Schwarzschild geometry in the asymptotic region,
\begin{align}
    F(r)\longrightarrow 1-\frac{2M}{r},\qquad r\rightarrow\infty,
\end{align}
while the spacetime is regular at $r\rightarrow0$. It is convenient to introduce the dimensionless quantities
\begin{align}
\label{coord}
    \tilde{g}\equiv\frac{g}{M}, \qquad \tilde r\equiv\frac{r}{M},\qquad \tilde{\alpha}\equiv M^3\alpha.
\end{align}
Hereafter, for notational simplicity, we omit the tildes and use $g$, $r$, and $\alpha$ to denote the corresponding dimensionless quantities.

For null geodesics restricted to the equatorial plane, the conserved energy and angular momentum are
\begin{align}
    E=F(r)\dot{t},\qquad L=r^2\dot{\phi},
\end{align}
and the radial equation can be written as
\begin{align}
    \frac{1}{2}e^{-\alpha r^3}\dot r^2+V_{\rm null}(r)
    =
    \frac{1}{2}E^2,
\end{align}
with the effective potential
\begin{align}
    V_{\rm null}(r)
    =
    \frac{L^2}{2r^2}F(r).
\end{align}
The impact parameter is defined by
\begin{align}
    u\equiv\frac{L}{E}.
\end{align}

The photon sphere corresponds to an unstable circular null orbit and is determined by the condition
\begin{align}
    \frac{dV_{\rm null}}{dr}=0.
\end{align}
Equivalently,
\begin{align}
    rF'(r)-2F(r)=0.
\end{align}
For the Hayward metric, this condition becomes
\begin{align}
    r^6-3r^5+2g^3r^3+g^6=0.
\end{align}
To determine the range of $g$ for which a positive real solution $r>0$ exists, we regard the equation as a quadratic equation in $g^3$. Solving for $g^3$, we obtain
\begin{align}
    g^3= -r^3+\sqrt{3r^5} = r^{5/2}\left(\sqrt{3}-\sqrt{r}\right),
\end{align}
where the other branch gives $g^3<0$ and is therefore discarded for $g>0$.

The maximum value of $g$ is obtained by extremizing the right-hand side with respect to $r$:
\begin{align}
    \frac{d g^3}{dr}
    =
    \frac{r^{3/2}}{2}
    \left(5\sqrt{3}-6\sqrt{r}\right)
    =0.
\end{align}
Thus, the critical radius is
\begin{align}
r_{\rm P}=\frac{25}{12}.
\end{align}
Substituting this value into $g^3(r)$ gives
\begin{align}
g_{\rm P}^3=\frac{1}{2\sqrt{3}}\left(\frac{25}{12}\right)^{5/2}\simeq 1.8085,
\end{align}
and hence
\begin{align}
g_{\rm P} = \left[ \frac{1}{2\sqrt{3}}\left(\frac{25}{12}\right)^{5/2}\right]^{1/3}\simeq 1.2183.
\end{align}
Therefore, a positive real photon-sphere solution exists for
\begin{align}
\boxed{0<g\leq g_{\rm P}\simeq0.1.2183}.
\end{align}
At the critical value $g=g_{\rm P}$, the two positive photon-sphere
solutions merge at
\begin{align}
r_{\rm P}=\frac{25}{12}.
\end{align}

Let $r_P$ denote the radius of the unstable photon sphere. The
corresponding critical impact parameter, or shadow radius, is given by
\begin{align}
u_{\rm P}=\frac{r_{\rm P}}{\sqrt{F(r_{\rm P})}}.
\end{align}
This critical impact parameter separates photon trajectories that are captured from those that are scattered back to the observer.

To calculate the deflection angle, we use
\begin{align}
    \frac{d\phi}{dr}=\dfrac{e^{-\frac{\alpha}{2}r^3}}{r^2}\left(\frac{1}{u^2}-\frac{F(r)}{r^2}\right)^{-1/2}.
\end{align}
Let $r_0$ be the distance of closest approach. It is determined by
\begin{align}
    \frac{1}{u^2}=\frac{F(r_0)}{r_0^2}.
\end{align}
Introducing the inverse radial coordinate
\begin{align}
    v &= \frac{1}{r},
\end{align}
the total deflection angle is given by
\begin{align}
    \Delta\phi(v)
    &=
    2\int_0^{1/r_0}
    \dfrac{e^{-\frac{\alpha}{v^3}} dv}
    {\sqrt{
        u^{-2}-v^2F(1/v)
    }}
    -\pi.
\end{align}

\subsubsection{Weak-deflection limit}

For $M/r_0\ll1$ and $\alpha=0$, the deflection angle can be expanded as
\begin{align}
    \Delta\phi
    ={}&
    \frac{4M}{r_0}
    +\frac{15\pi-16}{4}
    \left(\frac{M}{r_0}\right)^2
    +\frac{244-45\pi}{6}
    \left(\frac{M}{r_0}\right)^3
    + \left(
        -130+\frac{3465\pi}{64}
        -\frac{15\pi}{8}\frac{g^3}{M^3}
    \right)
    \left(\frac{M}{r_0}\right)^4\notag\\
&+\left(\frac{7783}{10}-\frac{3465\pi}{16}+\frac{75\pi-472}{10}\frac{g^3}{M^3}\right)\left(\frac{M}{r_0}\right)^5\notag\\
&+\left(\frac{310695\pi}{256}-\frac{21397}{6}+\frac{5}{32}(1664-693\pi)\frac{g^3}{M^3}\right)\left(\frac{M}{r_0}\right)^6+\mathcal{O}\left(\left(\frac{M}{r_0}\right)^7\right),
\end{align}
Alternatively, expressing the result in terms of the impact parameter
$b$, one obtains
\begin{align}
    \Delta\phi
    ={}&
    \frac{4M}{r_0}
    +\frac{15\pi}{4}
    \left(\frac{M}{u}\right)^2
    +\frac{128}{3}
    \left(\frac{M}{u}\right)^3
    + \left(\frac{3465\pi}{64}
        -\frac{15\pi}{8}\frac{g^3}{M^3}
    \right)
    \left(\frac{M}{u}\right)^4\notag\\
&+\left(\frac{3584}{5}-\frac{512}{10}\frac{g^3}{M^3}\right)\left(\frac{M}{u}\right)^5+\left(\frac{255255\pi}{256}-\frac{3465}{32}\frac{g^3}{M^3}\right)\left(\frac{M}{u}\right)^6+\mathcal{O}\left(\left(\frac{M}{u}\right)^7\right).
\end{align}
Therefore, the deviation from the Schwarzschild deflection angle is suppressed in the weak-field regime.

\subsubsection{Strong-deflection behavior}

The behavior of the deflection angle changes drastically when the
impact parameter approaches the critical value $u_{\rm P}$. For
\begin{equation}
    u\rightarrow u_{\rm P},
\end{equation}
the photon spends an increasingly long time in the vicinity of the
unstable photon sphere. Consequently, the deflection angle diverges
logarithmically,
\begin{equation}
    \Delta\phi
    \sim
    \frac{e^{-\alpha r_{\rm P}^3}}
    {\sqrt{-5+2\sqrt{3r_P/M}}}
    \ln(u-u_P)^{-1}.
\end{equation}
Thus, the photon sphere determines the critical behavior of the
strong-deflection limit.

At the critical value
\begin{equation}
    a=a_{\rm P},
\end{equation}
the photon sphere becomes marginal and the nature of the divergence
changes. In this case,
\begin{equation}
    r_{\rm P}=\frac{25}{12}M,
\end{equation}
and the deflection angle behaves as
\begin{equation}
    \Delta\phi
    \sim
    c_0 M^{1/6}(b-b_{\rm P})^{-1/6},
\end{equation}
where
\begin{equation}
    c_0\simeq 7.771.
\end{equation}
Therefore, the usual logarithmic divergence associated with an unstable photon sphere is replaced by a power-law divergence at the critical value $a=a_{\rm P}$.

\subsubsection{Light trajectories and image formation}

The parameter $g/M$ also changes the qualitative behavior of photon trajectories. For $g/M<(g/M)_{\rm P}$, an unstable photon sphere exists, and photons with impact parameters close to the critical value $u_{\rm P}$ can wind around the central object several times before reaching a distant observer.

For example, when $g/M=0.5$, the resulting image is qualitatively similar to that of the Schwarzschild black hole, with a bright ring surrounding a central dark region associated with the black-hole shadow.

A more interesting situation occurs in the parameter range
\begin{equation}
\left(\frac{g}{M}\right)_{\rm H}
<\frac{g}{M}
<\left(\frac{g}{M}\right)_{\rm P},
\end{equation}
where $(g/M)_{\rm H}$ denotes the critical value above which the event horizon ceases to exist,
\begin{equation}
\left(\frac{g}{M}\right)_{\rm H}
=\frac{2^{5/3}}{3}
\simeq 0.83995.
\end{equation}
In this parameter range, the event horizon is absent while the unstable photon sphere still exists. Consequently, the central dark region is no longer associated with a black-hole shadow and can instead exhibit a doughnut-shaped structure, while a bright ring remains.

For
\begin{equation}
\frac{g}{M}
>\left(\frac{g}{M}\right)_{\rm P},
\end{equation}
the photon sphere no longer exists. Nevertheless, when $g/M$ is only slightly larger than $(g/M)_{\rm P}$, the deflection angle can still exceed $2\pi$. Photons can therefore wind around the central region more than once, producing a bright ring even in the absence of an unstable photon sphere.

This result demonstrates that the presence of a bright ring in an observed image does not necessarily imply the existence of an unstable photon sphere. Rather, the appearance of such a ring is more generally determined by the behavior of null geodesics and the large deflection of light in the vicinity of the critical parameter region.

In this section, we have reviewed the gravitational lensing of photons in the Hayward black hole spacetime without EFT corrections. We have derived the photon-sphere radius, the corresponding critical impact parameter, and the strong-deflection expansion of the deflection angle, and have examined how the Hayward parameter affects the resulting lensing observables. These results provide the baseline predictions for the Hayward spacetime against which the effects of EFT corrections can be assessed.

In the following section, we extend this analysis by incorporating the curvature-dependent EFT corrections to photon propagation. In particular, non-minimal couplings between the electromagnetic field and the spacetime curvature modify the photon dispersion relation and can lead to polarization-dependent propagation. This effect is especially relevant for the Hayward spacetime because, unlike the Schwarzschild vacuum, it is generally non-Ricci-flat, $R_{\mu\nu}\neq0$. Consequently, the Ricci-photon coupling $R_{\mu\nu}F^{\mu\rho}F^{\nu}_{\ \rho}$ can contribute to photon propagation in the Hayward background, whereas it vanishes identically in the Schwarzschild spacetime. The strong-deflection regime, in which the deflection angle develops a logarithmic divergence near the critical impact parameter, may therefore provide enhanced sensitivity to these perturbatively small EFT corrections. In the next section, we investigate these effects by deriving the modified photon propagation law and the corresponding effective optical metrics, and then examine their impact on the strong-deflection lensing observables.

\section{Strong-Deflection Limit in EFT-Corrected Hayward black hole}
\label{sec:strong lens}

In the previous section, we investigated the strong-deflection limit of light propagation in the Hayward black hole spacetime in the absence of EFT corrections, providing the baseline lensing observables for comparison. In this section, we incorporate EFT corrections to photon propagation in the Hayward background. These curvature-dependent corrections modify the photon propagation law and lead to polarization-dependent effective optical metrics for the physical photon modes. We investigate how these modifications affect the photon-sphere radius, critical impact parameter, and strong-deflection observables, and assess whether the combined effects of the Hayward geometry and EFT-corrected photon propagation can provide characteristic signatures for distinguishing the Hayward regular black hole from the Schwarzschild black hole.

We consider a general static and spherically symmetric spacetime and restrict the photon motion to the equatorial plane, $\theta=\pi/2$. The metric can then be written as
\begin{align}
ds^2=-A(r)dt^2+B(r)dr^2+C(r)d\phi^2,
\end{align}
where $A(r)$, $B(r)$, and $C(r)$ are functions of the radial coordinate. We denote by $r_0$ the radius of closest approach of a photon trajectory and define $A_0=A(r_0)$ and $C_0=C(r_0)$. Following the strong-deflection formalism developed by Bozza~\cite{Bozza:2002zj,Bozza:2002af,Bozza:2003cp}, we introduce the variable
\begin{align}
\label{zcoor}
z=1-\frac{r_0}{r}.
\end{align}
The total change in the azimuthal angle along the photon trajectory can then be expressed as
\begin{align}
\Delta\phi\equiv I(r_0)
=\int_0^1 R(z,r_0)f(z,r_0),dz,
\end{align}
where
\begin{align}
R(z,r_0)
&=\frac{2r_0\sqrt{A(r(z))B(r(z))C_0}}{C(r(z))(1-z)^2},
\\
f(z,r_0)
&=\frac{1}{\sqrt{A_0-A(r(z))C_0/C(r(z))}}.
\end{align}
Here and below, a prime denotes differentiation with respect to $r$, and quantities carrying the subscript $0$ are evaluated at $r=r_0$.

The function $R(z,r_0)$ is regular in the relevant region, whereas the divergence of the integral originates from $f(z,r_0)$ as $z\rightarrow0$. To isolate the leading singular behavior, we expand the argument of the square root around $z=0$ and retain terms up to second order:
\begin{align}
f(z,r_0)\simeq f_0(z,r_0)
=\frac{1}{\sqrt{p(r_0)z+q(r_0)z^2}},
\label{pole expansion}
\end{align}
where
\begin{align}
p(r_0)
&=\frac{r_0}{C_0}\left(C_0'A_0-C_0A_0'\right),
\end{align}
and $q(r_0)$ is determined by the corresponding second-order expansion. When $p(r_0)\neq0$, the integrand behaves as $z^{-1/2}$ near $z=0$ and is therefore integrable. The deflection angle becomes divergent when $p(r_0)=0$, for which the leading behavior changes to $z^{-1}$.

The condition $p(r_{\rm ph})=0$ determines the radius of the unstable circular photon orbit, or photon sphere,
\begin{align}
\left.
\frac{d}{dr}\left(\frac{C(r)}{A(r)}\right)
\right|_{r=r_{\rm ph}}=0.
\label{ph radius}
\end{align}
The corresponding critical impact parameter is
\begin{align}
u_{\rm ph}
=\sqrt{\frac{C_{\rm ph}}{A_{\rm ph}}},
\label{critical impact}
\end{align}
where $A_{\rm ph}=A(r_{\rm ph})$ and $C_{\rm ph}=C(r_{\rm ph})$. As the closest approach radius approaches the photon-sphere radius from the appropriate side, $r_0\rightarrow r_{\rm ph}$, the photon spends an increasingly long time in the vicinity of the unstable circular orbit, resulting in a logarithmic divergence of the deflection angle.

To separate the divergent and regular contributions, we decompose the integral as
\begin{align}
I(r_0)=I_D(r_0)+I_R(r_0),
\end{align}
where
\begin{align}
I_D(r_0)
&=\int_0^1R(0,r_{\rm ph})f_0(z,r_0),dz,\\
I_R(r_0)
&=\int_0^1g(z,r_0),dz,
\label{regular angle}
\end{align}
with
\begin{align}
g(z,r_0)
=R(z,r_0)f(z,r_0)
-R(0,r_{\rm ph})f_0(z,r_0).
\end{align}
By construction, $I_D$ contains the leading divergent behavior, while $I_R$ remains finite in the limit $r_0\rightarrow r_{\rm ph}$.

The divergent contribution can be evaluated analytically as
\begin{align}
I_D(r_0)=
\frac{2R(0,r_{\rm ph})}{\sqrt{q(r_0)}}
\log\left[
\frac{\sqrt{q(r_0)}+\sqrt{p(r_0)+q(r_0)}}
{\sqrt{p(r_0)}}
\right].
\end{align}
Expanding this expression around the photon-sphere radius gives
\begin{align}
I_D(r_0)=-a\log\left(\frac{r_0}{r_{\rm ph}}-1\right)
+b_D
+\mathcal{O}!\left[
(r_0-r_{\rm ph})\log(r_0-r_{\rm ph})
\right],
\end{align}
where
\begin{align}
a&=\frac{R(0,r_{\rm ph})}{\sqrt{q(r_{\rm ph})}},\\
b_D&=\frac{R(0,r_{\rm ph})}{\sqrt{q(r_{\rm ph})}}\log2.
\end{align}
Thus, the leading divergence of the deflection angle is logarithmic as the closest approach radius approaches the photon sphere.

The regular contribution is obtained by evaluating $I_R$ at the photon sphere. Since the subtraction in Eq.~\eqref{regular angle} removes the singular part of the integrand, the resulting function $g(z,r_{\rm ph})$ is regular over the integration domain. We therefore obtain
\begin{align}
I_R(r_0)=\int_0^1g(z,r_{\rm ph}),dz
+\mathcal{O}(r_0-r_{\rm ph}),
\end{align}
and define
\begin{align}
b_R=I_R(r_{\rm ph})
=\int_0^1g(z,r_{\rm ph}),dz.
\end{align}
Combining the divergent and regular contributions, the total deflection angle is written as
\begin{align}
I(r_0)=-a\log\left(\frac{r_0}{r_{\rm ph}}-1\right)
+b
+\mathcal{O}!\left[
(r_0-r_{\rm ph})\log(r_0-r_{\rm ph})
\right],
\end{align}
where
\begin{align}
b=-\pi+b_D+b_R.
\end{align}

To express the strong-deflection expansion in terms of the impact parameter, we introduce
\begin{align}
u=\sqrt{\frac{C_0}{A_0}}.
\label{impact para}
\end{align}
At the photon sphere, this reduces to the critical impact parameter $u_{\rm ph}$. Since $du/dr_0$ vanishes at $r_0=r_{\rm ph}$, the expansion of $u$ around the critical orbit begins at quadratic order,
\begin{align}
\label{c_def}
u-u_{\rm ph}=c(r_0-r_{\rm ph})^2
+\mathcal{O}!\left((r_0-r_{\rm ph})^3\right),
\end{align}
where $c$ is determined by the metric functions and their derivatives evaluated at the photon sphere.

Using the relation between the impact parameter and the angular position of the image, $u\simeq D_{OL}\theta$, the deflection angle in the strong-deflection limit takes the standard form
\begin{align}
\alpha(\theta)=-\bar a
\log\left(
\frac{\theta D_{OL}}{u_{\rm ph}}-1
\right)
+\bar b,
\end{align}
where
\begin{align}
\bar a
&=\frac{a}{2}
=\frac{R(0,r_{\rm ph})}{2\sqrt{q(r_{\rm ph})}},
\\
\bar b
&=
b+\frac{a}{2}
\log\left(
\frac{cr_{\rm ph}^2}{u_{\rm ph}}
\right)
\notag\\
&=
-\pi+b_R
+\bar a\log\left(
\frac{2q(r_{\rm ph})}{A_{\rm ph}}
\right).
\end{align}
The coefficients $\bar a$ and $\bar b$, together with the critical impact parameter $u_{\rm ph}$, characterize the leading strong-deflection behavior and determine the corresponding relativistic lensing observables.

In the following section, we apply these general results to the Hayward black hole spacetime. We first derive the photon-sphere radius, critical impact parameter, and strong-deflection coefficients in the absence of EFT corrections. These results provide the reference values against which the modifications induced by the curvature-dependent photon propagation are subsequently evaluated.

\subsection{Strong Deflection Analysis of EFT-corrected Hayward black hole}

In this subsection, we investigate gravitational lensing in the strong-deflection regime, taking into account the effects of the curvature-photon couplings introduced in Eq.~\eqref{eq:action}. We consider the Hayward regular black hole as the background spacetime and focus on the modifications to photon propagation induced by the EFT corrections. Throughout this analysis, we adopt dimensionless coordinates and parameters by measuring all lengths in units of the black-hole mass, and set $M=1$ without loss of generality. The remaining dimensionless parameters therefore characterize the deviation of the Hayward geometry from the Schwarzschild spacetime.

For photons propagating in the equatorial plane, the EFT-corrected propagation law derived from the curvature-photon interactions can be recast in terms of effective optical metrics. The two physical polarization modes, PPL and PPM, experience different effective geometries as a consequence of the Weyl- and Ricci-curvature couplings. Using Eq.~(\ref{metric_1}), which is valid to first order in the EFT coupling constants, the corresponding effective metrics can be written in a unified form as
\begin{align}
\label{metric_2}
ds^2=&-\left(1-\dfrac{2Mr^2}{r^3+g^3}\right)\left(1+\beta\mathcal{C}\right)^{-1}dt^2+\dfrac{1}{\left(1-\dfrac{2Mr^2}{r^3+g^3}\right)}\left(1+\beta\mathcal{D}\right)^{-1}dr^2\notag\\
&+r^2\left(1+\beta\mathcal{E}\right)^{-1}\left(\dfrac{1+8\gamma\mathcal{A}}{1+8\gamma\mathcal{B}}\right)^s(d\theta^2+\sin^2\theta d\phi^2),\\
\mathcal{A}=&-\dfrac{2M(r^6-3Mg^3r^3+2Mg^6)}{\left(g^{3}+r^{3}\right)^{3}},\\
\mathcal{B}=&\dfrac{M(r^6-4g^6)}{\left(g^{3}+r^{3}\right)^{3}},\\
\mathcal{C}=&-\dfrac{12Mg^3(g^3-2r^3)}{(r^3+g^3)^3},\\
\mathcal{D}=&\dfrac{12Mg^3(2r^3-g^3)}{(r^3+g^3)^3},\\
\mathcal{E}=&-\dfrac{12Mg^3}{(r^3+g^3)^2}.
\end{align}
In this paper, we perform our analysis using the dimensionless coordinates introduced in Eq.~\eqref{coord}. In these coordinates, the metric above takes the form obtained by setting $M=1$.

The strong-deflection analysis then allows us to derive the lensing observables while retaining the dependence on the physical parameters. In particular, the deflection angle is decomposed into a logarithmically divergent part and a regular contribution. The divergent part can be obtained analytically. For $g^3=0$, corresponding to the Schwarzschild limit, the regular contribution $b_R$ can also be evaluated analytically. For the nonzero values of $g^3$ considered here, however, the integral defining $b_R$ cannot be evaluated in closed form. We therefore evaluate $b_R$ numerically for the corresponding parameter choices.

For the spacetime under consideration, the photon-sphere radius, including the leading EFT corrections, is given for each value of $g^3$ in Table~\ref{tab:photon_sphere}.
\begin{table}[t]
\centering
\caption{Photon-sphere radius $r_{\rm ph}$ for different values of $g^3$, with $M=1$.}
\label{tab:photon_sphere}
\begin{tabular}{c|c|c}
\hline\hline
$g^3$ & Numerical value of $g^3$ & $r_{\rm ph}$ \\
\hline
$0$ & 0 & $3-\frac{4s\gamma}{3}$ \\
\hline
$\left(\frac{119}{40}\right)^{5/2}\left(\sqrt{3}-2\sqrt{\dfrac{119}{40}}\right)$ & 0.1104 & $\frac{119}{40}+\frac{20(-284529\beta+4764\sqrt{3570}\beta-502418s\gamma+8256\sqrt{3570}s\gamma)}{14161(-50+\sqrt{3570})}$\\
\hline
$\frac{1}{3}\left(\frac{8}{3}\right)^{5/2}(3\sqrt{3}-2\sqrt{6})$ & 1.1503 & $\frac{8}{3}+\frac{9(27-19\sqrt{2})\beta+60(7-5\sqrt{2})s\gamma)}{32-20\sqrt{2}}$ \\
\hline
$\frac{1}{3}\left(\frac{7}{3}\right)^{5/2}(3\sqrt{3}-\sqrt{21})$ & 1.7972 & $\frac{7}{3}+\frac{12(9(8-3\sqrt{7})\beta+(124-47\sqrt{7})s\gamma)}{98-35\sqrt{7}}$ \\
\hline
$\frac{1}{6}\left(\frac{13}{6}\right)^{5/2}(6\sqrt{3}-\sqrt{78})$ & 1.8085 & $\frac{13}{6}-\frac{6}{169}(15(-13+\sqrt{26})\beta+2(-117+23\sqrt{26})s\gamma)$ \\
\hline\hline
\end{tabular}
\end{table}
The photon-sphere radius plays a central role in determining the properties of strong gravitational lensing. In particular, it determines the critical impact parameter $u_c$, which characterizes the boundary between photons that are scattered back to the observer and those that are captured by the central object. The critical impact parameter is directly related to the angular positions of the relativistic images and therefore provides a natural connection between the photon-sphere structure and observable lensing quantities. We thus use the photon-sphere radius obtained above as a key ingredient in the subsequent analysis of the strong-deflection limit.

In this background, we analyze the deflection angle in the strong-deflection limit following the standard formalism. We first consider the $g^3=0$ case, which corresponds to the Schwarzschild limit. In this case, the relevant functions $R(z,r_{\rm ph})$ and $f(z,r_{\rm ph})$ appearing in the integral expression for the deflection angle are given by
\begin{align}
R(z,r_{\rm ph})=&2-\frac{8}{9}s\gamma(1-2(1-z)^3),\\
f(z,r_{\rm ph})=&\dfrac{1}{\sqrt{z^2-\frac{2z^3}{3}}}+\dfrac{4s\gamma z(3-3z+z^2)}{9\sqrt{z^2-\frac{2z^3}{3}}},
\end{align}
where $z$ is defined in terms of the radial coordinate $r$.

For $g^3=0$, the integral for the regular contribution $b_R$ can be evaluated analytically. We obtain
\begin{align}
b_R=-4\arctan\left(\frac{1}{\sqrt{3}}\right)+\log36-\frac{8s\gamma}{45}(18-4\sqrt{3}-5\log12+10\log(1+\sqrt{3})).
\end{align}

For the nonzero values of $g^3$ considered below, the integral defining the regular contribution $b_R$ does not admit a closed-form expression. Nevertheless, the strong-deflection coefficient $\bar a$, and the critical impact parameter $u_{\rm ph}$ can be obtained analytically for each value of $g^3$. We therefore evaluate only the regular contribution $b_R$ numerically for the nonzero values of $g^3$, while $\bar a$, and $u_{\rm ph}$ are determined analytically. The resulting expressions for $\bar a$, and $u_{\rm ph}$ are summarized in Tables~\ref{tab:bara}--\ref{tab:uph}.
 
\begin{table}[htbp]
\centering
\caption{Strong-deflection coefficient $\bar{a}$ for different values of $g^3$, including the leading-order EFT corrections, with $M=1$.}
\label{tab:bara}
\begin{tabular}{c|c}
\hline
$g^3$ & $\bar{a}$ \\
\hline
$0$ & $1+\dfrac{4s\gamma}{9}$\\
\hline
$\left(\frac{119}{40}\right)^{5/2}\left(\sqrt{3}-2\sqrt{\dfrac{119}{40}}\right)$ & $\sqrt{\dfrac{10}{-50+\sqrt{3570}}}-\dfrac{200\sqrt{10}(-1634907\beta+27364\sqrt{3570}\beta-3368534s\gamma+56328\sqrt{3570}s\gamma)}{14161(-50+\sqrt{3570})^{5/2}}$ \\
\hline
 $\frac{1}{3}\left(\frac{8}{3}\right)^{5/2}(3\sqrt{3}-2\sqrt{6})$ & $\dfrac{2}{\sqrt{-5+4\sqrt{2}}}-\dfrac{9((-899+636\sqrt{2})\beta+32(-58+41\sqrt{2})s\gamma)}{16(-5+4\sqrt{2})^{5/2}}$ \\
\hline
 $\frac{1}{3}\left(\frac{7}{3}\right)^{5/2}(3\sqrt{3}-\sqrt{21})$ & $\dfrac{1}{\sqrt{-5+2\sqrt{7}}}-\dfrac{36(2(-511+193\sqrt{7})\beta+(-2114+799\sqrt{7})s\gamma)}{49(-5+2\sqrt{7})^{3/2}(-14+5\sqrt{7})}$ \\
\hline
 $\frac{1}{6}\left(\frac{13}{6}\right)^{5/2}(6\sqrt{3}-\sqrt{78})$ & $\dfrac{1}{\sqrt{-5+\sqrt{26}}}+\dfrac{18((-293+53\sqrt{26})\beta+2(-293+57\sqrt{26})s\gamma)}{169(-5+\sqrt{26})^{3/2}}$ \\
\hline
\end{tabular}
\end{table}

\begin{table}[htbp]
\centering
\caption{Critical impact parameter $u_{\rm ph}$ for different values of $g^3$, including the leading-order EFT corrections, with $M=1$.}
\label{tab:uph}
\begin{tabular}{c|c}
\hline
$g^3$ & $u_{\rm ph}$ \\
\hline
$0$ & $3\sqrt{3}-\dfrac{4s\gamma}{\sqrt{3}}$\\
\hline
$\left(\frac{119}{40}\right)^{5/2}\left(\sqrt{3}-2\sqrt{\dfrac{119}{40}}\right)$ & $\frac{119}{40}\sqrt{\dfrac{357}{357-4\sqrt{3570}}}+40\sqrt{2}\dfrac{(-391629+6555\sqrt{3570})\beta+2(-327369+5441\sqrt{3570})s\gamma}{\sqrt{42483-476\sqrt{3570}}(-32130+557\sqrt{3570})}$ \\
\hline
 $\frac{1}{3}\left(\frac{8}{3}\right)(3\sqrt{3}-2\sqrt{6})^{5/2}$ & $\dfrac{8\sqrt{2+\sqrt{2}}}{3}-\dfrac{3((-75+53\sqrt{2})\beta+4(-31+22\sqrt{2})s\gamma)}{(26-18\sqrt{2})\sqrt{4-2\sqrt{2}}}$ \\
\hline
 $\frac{1}{3}\left(\frac{7}{3}\right)(3\sqrt{3}-\sqrt{21})^{5/2}$ & $\dfrac{7\cdot7^{1/4}}{3\sqrt{-2+\sqrt{7}}}+\dfrac{18((-463+175\sqrt{7})\beta+6(-127+48\sqrt{7})s\gamma)}{7^{1/4}(-2+\sqrt{7})^{3/2}(-140+53\sqrt{7})}$ \\
\hline
 $\frac{1}{6}\left(\frac{13}{6}\right)(6\sqrt{3}-\sqrt{78})^{5/2}$ & $\frac{13}{6}\sqrt{\frac{1}{5}(13+2\sqrt{26})}+\dfrac{6(5(-26+5\sqrt{26})\beta+2(-104+21\sqrt{26})s\gamma)}{\sqrt{169-26\sqrt{26}}(-13+2\sqrt{26})}$ \\
\hline
\end{tabular}
\end{table}

The regular contribution $b_R$ and the resulting strong-deflection coefficient $\bar b$ are summarized in Table~\ref{tab:b}. For $g^3=0$, both $b_R$ and $\bar b$ are obtained analytically, whereas for $g^3\neq0$, $b_R$ is evaluated numerically and $\bar b$ is subsequently determined from the numerical result.
\begin{table}[htbp]
\centering
\caption{Numerical values of the regular contribution $b_R$ and the strong-deflection coefficient $\bar{b}$ for different values of $g^3$, with the background contribution $b_R^{(0)}$ and the EFT corrections $b_R^{(\beta)}$, $b_R^{(\gamma)}$ proportional to $\beta$ and $s\gamma$ shown separately, with $M=1$.}
\label{tab:b}
\begin{tabular}{c|c|c|c|c}
\hline
$g^3$ & $b_R^{(0)}$ & $b_R^{(\beta)}$ & $b_R^{(\gamma)}$ & $\bar{b}$\\
\hline
$0$ & $0.949603$ & $0$ & $-1.54627$ & $1.39153+1.82418s\gamma$\\
\hline
$\left(\frac{119}{40}\right)^{5/2}\left(\sqrt{3}-2\sqrt{\dfrac{119}{40}}\right)$ & 0.930524 & 0.26966 & -1.55516 &$-1.25107+0.25469\beta+0.14844s\gamma$ \\
\hline
 $\frac{1}{3}\left(\frac{8}{3}\right)(3\sqrt{3}-2\sqrt{6})^{5/2}$ & 0.537146 & 0.85577 & -1.63188 & $-0.75267+0.62203\beta+0.358455s\gamma$\\
\hline
 $\frac{1}{3}\left(\frac{7}{3}\right)(3\sqrt{3}-\sqrt{21})^{5/2}$ & -1.17032 & 15.8812 & 0.383285 & $3.58402+18.4643\beta+3.68244s\gamma$\\
\hline
 $\frac{1}{6}\left(\frac{13}{6}\right)(6\sqrt{3}-\sqrt{78})^{5/2}$ & -7.19614 & 426.72139 & 89.33366 & $-10.6068+575.287\beta-52.7243s\gamma$\\
\hline
\end{tabular}
\end{table}

Combining the divergent and regular contributions, the strong-deflection coefficient $\bar b$ for the $g^3=0$ case is obtained analytically as
\begin{align}
\bar{b}=-\pi-4\arctan\left(\frac{1}{\sqrt{3}}\right)+\log1296+\frac{8}{45}s\gamma\left(-8+4\sqrt{3}+\log1934917632-10\log(1+\sqrt{3})\right).
\end{align}
For $g^3\neq0$, the corresponding value of $\bar b$ is obtained from the numerical evaluation of the divergent and regular contributions, as summarized in Table~\ref{tab:b}.

Finally, the deflection angle in the strong-deflection limit can be written in terms of the angular position $\theta$ as
\begin{align}
\alpha(\theta)=&-\left(1+\frac{4s\gamma}{9}\right)\log\left(\frac{\theta D_{OL}}{3\sqrt{3}-\frac{4s\gamma}{\sqrt{3}}} - 1 \right)+\bar{b}+\mathcal{O}(u-u_{\rm ph}),
\end{align}
for the $g^3=0$ case. For the nonzero values of $g^3$, the strong-deflection angle is evaluated numerically using the corresponding values of $\bar a$, $\bar b$, and $u_{\rm ph}$ given above.

\begin{figure}[t]
    \centering

    \begin{subfigure}{0.32\textwidth}
        \centering
        \includegraphics[width=\textwidth]{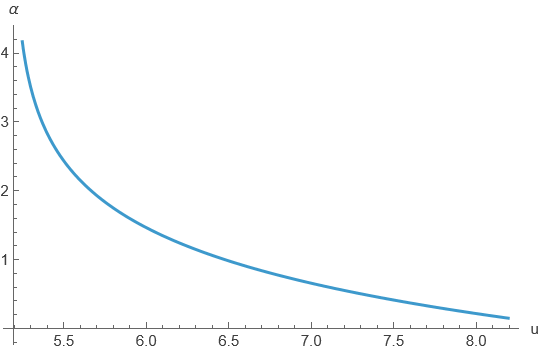}
        \caption{$g^3=0$}
    \end{subfigure}
    \hfill
    \begin{subfigure}{0.32\textwidth}
        \centering
        \includegraphics[width=\textwidth]{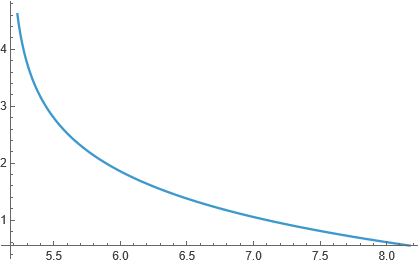}
        \caption{$g^3=\left(\frac{119}{40}\right)^{5/2}\left(\sqrt{3}-2\sqrt{\dfrac{119}{40}}\right)$}
    \end{subfigure}
    \hfill
    \begin{subfigure}{0.32\textwidth}
        \centering
        \includegraphics[width=\textwidth]{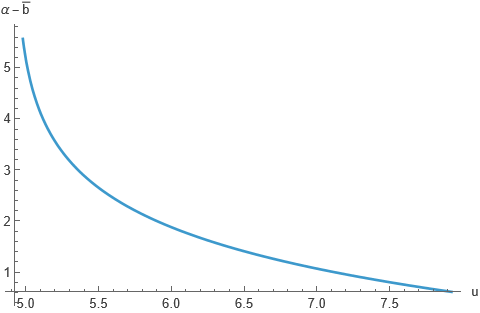}
        \caption{$g^3=\frac{1}{3}\left(\frac{8}{3}\right)^{5/2}(3\sqrt{3}-2\sqrt{6})$}
    \end{subfigure}
    \hfill
    \begin{subfigure}{0.32\textwidth}
        \centering
        \includegraphics[width=\textwidth]{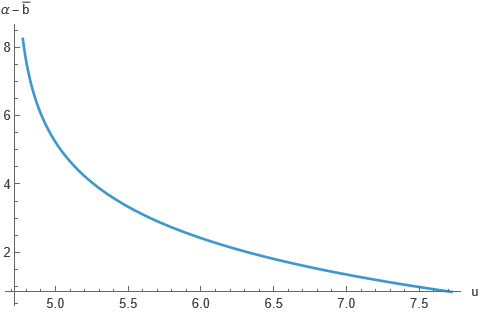}
        \caption{$g^3=\frac{1}{3}\left(\frac{7}{3}\right)^{5/2}(3\sqrt{3}-\sqrt{21})$}
    \end{subfigure}
    \hfill
    \begin{subfigure}{0.32\textwidth}
        \centering
        \includegraphics[width=\textwidth]{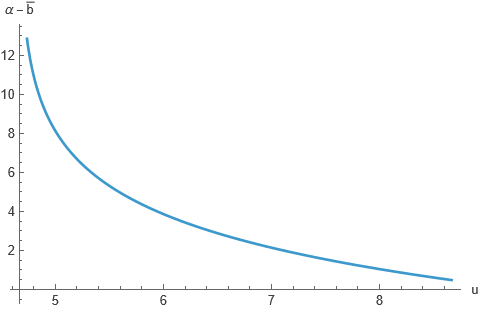}
        \caption{$g^3=\frac{1}{6}\left(\frac{13}{6}\right)^{5/2}(6\sqrt{3}-\sqrt{78})$}
    \end{subfigure}

    \caption{Deflection angle $\alpha(u)$ for the higher-curvature-corrected Hayward black hole with $s=1$ (PPL mode) and higher-curvature coupling constants $\beta=\gamma=10^{-3}$. For clarity, the constant term $\bar{b}$ is subtracted from the deflection angle for $g^3\neq0$, while the $g^3=0$ case is shown without subtracting $\bar{b}$.}
    \label{fig:RB0_deflection_angle}
\end{figure}

\begin{figure}[t]
    \centering

    \begin{subfigure}{0.32\textwidth}
        \centering
        \includegraphics[width=\textwidth]{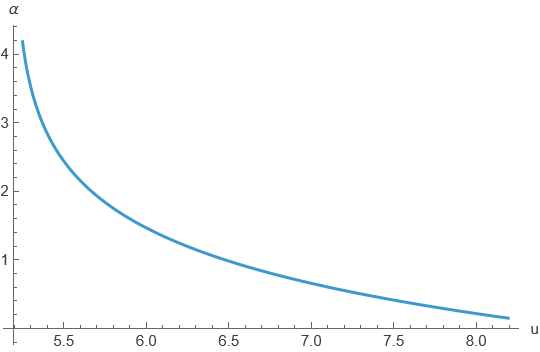}
        \caption{$g^3=0$}
    \end{subfigure}
    \hfill
\begin{subfigure}{0.32\textwidth}
        \centering
        \includegraphics[width=\textwidth]{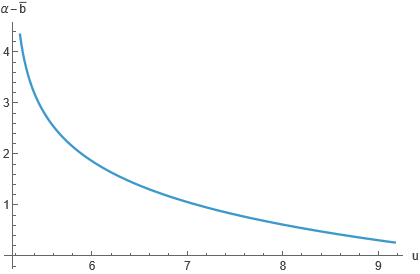}
        \caption{$g^3=\left(\frac{119}{40}\right)^{5/2}\left(\sqrt{3}-2\sqrt{\dfrac{119}{40}}\right)$}
    \end{subfigure}
    \hfill
    \begin{subfigure}{0.32\textwidth}
        \centering
        \includegraphics[width=\textwidth]{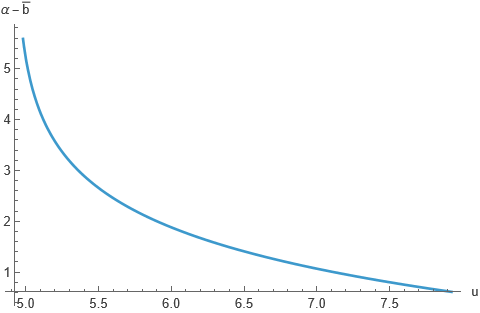}
        \caption{$g^3=\frac{1}{3}\left(\frac{8}{3}\right)^{5/2}(3\sqrt{3}-2\sqrt{6})$}
    \end{subfigure}
    \hfill
    \begin{subfigure}{0.32\textwidth}
        \centering
        \includegraphics[width=\textwidth]{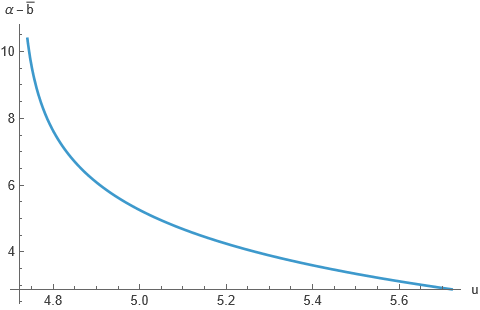}
        \caption{$g^3=\frac{1}{3}\left(\frac{7}{3}\right)^{5/2}(3\sqrt{3}-\sqrt{21})$}
    \end{subfigure}
    \hfill
    \begin{subfigure}{0.32\textwidth}
        \centering
        \includegraphics[width=\textwidth]{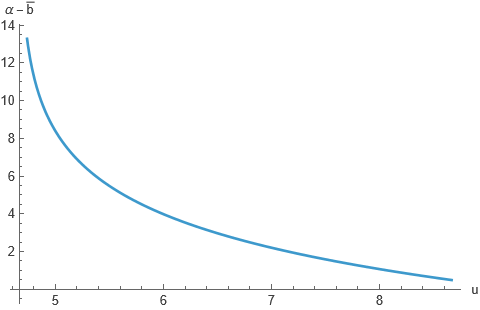}
        \caption{$g^3=\frac{1}{6}\left(\frac{13}{6}\right)^{5/2}(6\sqrt{3}-\sqrt{78})$}
    \end{subfigure}

    \caption{Deflection angle $\alpha(u)$ for the Hayward black hole without EFT corrections. For clarity, the constant term $\bar{b}$ is subtracted from the deflection angle for $g^3\neq0$, while the $g^3=0$ case is shown without subtracting $\bar{b}$.}
    \label{fig:RB00}
\end{figure}

For each of the five representative values of $g^3$ considered above, we compare the deflection angles with and without the EFT corrections. For $g^3=0$, corresponding to the Schwarzschild limit, the full strong-deflection angle can be obtained analytically and is used in the comparison. The corresponding result is shown in Fig.~\ref{fig:RB0_deflection_angle}. For the nonzero values of $g^3$, we focus on the logarithmically divergent part of the deflection angle, since the contribution of the perturbative EFT corrections to this term may become more pronounced in the strong-deflection regime. The corresponding results are shown in Fig.~\ref{fig:RB00}. The EFT coupling constants are fixed to $\beta=\gamma=10^{-3}$ in all plots.

\begin{figure}[t]
    \centering

  \begin{subfigure}{0.32\textwidth}
        \centering
        \includegraphics[width=\textwidth]{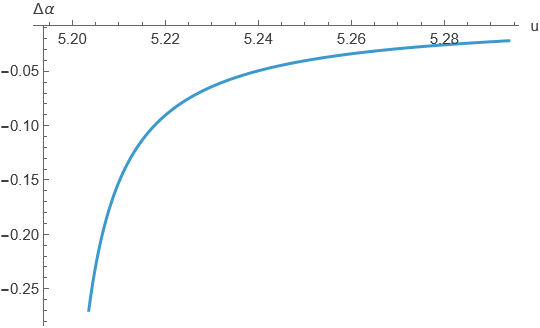}
        \caption{$g^3=0$}
    \end{subfigure}
    \hfill  
\begin{subfigure}{0.32\textwidth}
        \centering
        \includegraphics[width=\textwidth]{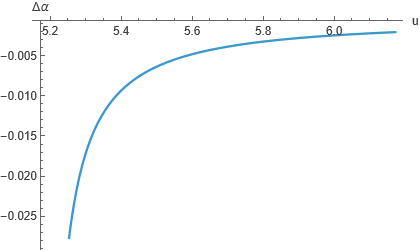}
        \caption{$g^3=\left(\frac{119}{40}\right)^{5/2}\left(\sqrt{3}-2\sqrt{\dfrac{119}{40}}\right)$}
    \end{subfigure}
    \hfill
  \begin{subfigure}{0.32\textwidth}
        \centering
        \includegraphics[width=\textwidth]{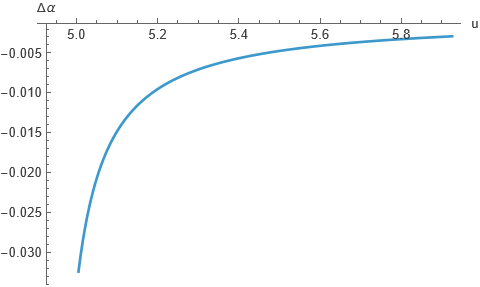}
        \caption{$g^3=\frac{1}{3}\left(\frac{8}{3}\right)^{5/2}(3\sqrt{3}-2\sqrt{6})$}
    \end{subfigure}
    \hfill
    \begin{subfigure}{0.32\textwidth}
        \centering
        \includegraphics[width=\textwidth]{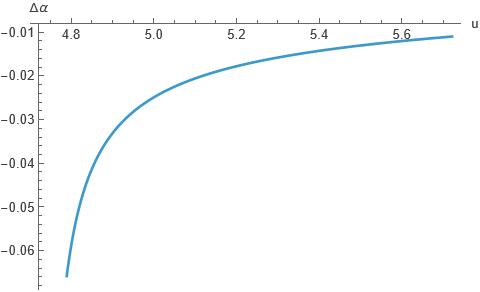}
        \caption{$g^3=\frac{1}{3}\left(\frac{7}{3}\right)^{5/2}(3\sqrt{3}-\sqrt{21})$}
    \end{subfigure}
    \hfill
    \begin{subfigure}{0.32\textwidth}
        \centering
        \includegraphics[width=\textwidth]{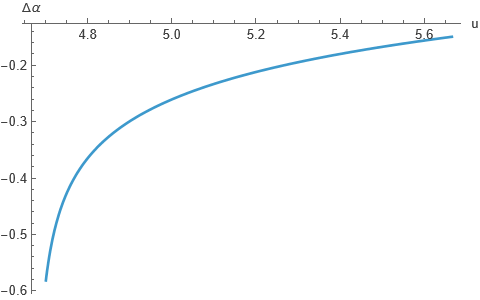}
        \caption{$g^3=\frac{1}{6}\left(\frac{13}{6}\right)^{5/2}(6\sqrt{3}-\sqrt{78})$}
    \end{subfigure}

    \caption{Difference in the deflection angle $\Delta\alpha$ between the higher-curvature-corrected Hayward black hole and the Hayward black hole without EFT corrections, with $s=1$ (PPL mode) and higher-curvature coupling constants $\beta=\gamma=10^{-3}$. For clarity, the constant term $\bar{b}$ is subtracted from the difference for $g^3\neq0$, while the $g^3=0$ case is shown without subtracting $\bar{b}$.}
    \label{fig:RB_deflection_delta}
\end{figure}

For each value of $g^3$, we also consider the difference between the deflection angles with and without the EFT corrections,
\begin{align}
\Delta\alpha = \alpha_{\rm EFT}-\alpha_{\rm 0},
\end{align}
where $\alpha_{\rm EFT}$ and $\alpha_{\rm 0}$ denote the deflection angles with and without the EFT corrections, respectively. The difference $\Delta\alpha$ isolates the effect of the EFT corrections at each fixed value of $g^3$. The corresponding results are shown in Fig.~\ref{fig:RB_deflection_delta}, allowing us to assess the contribution of the EFT corrections to the strong-deflection behavior.


The strong deflection limit corresponds to the case in which the closest approach $r_0$ approaches the radius of the photon sphere $r_{\rm ph}$. In this limit, the deflection angle exhibits a logarithmic divergence, which can be systematically analyzed using the formalism developed in the previous subsection. As demonstrated for the Hayward black hole in this section, even when the EFT corrections are parametrically small, their contributions to the logarithmically divergent term can become appreciable in the strong deflection regime. This enhancement provides a potentially sensitive probe of higher-curvature and higher-order electromagnetic interactions in the near-photon-sphere region.

One of the main goals of this work is to identify observational signatures that can distinguish the Hayward black hole from the Schwarzschild black hole through gravitational lensing. In particular, the Hayward black hole reduces to the Schwarzschild spacetime in the limit $g^3=0$, while nonzero values of $g^3$ modify the geometry in the near-photon-sphere region. These geometric modifications, together with the EFT corrections, affect the photon-sphere radius, the critical impact parameter, and the coefficients governing the logarithmically divergent part of the deflection angle. Although the EFT corrections may be parametrically small, their contributions to the strong-deflection coefficients can become appreciable in the strong deflection regime. The resulting modifications of the lensing observables may therefore provide characteristic signatures that distinguish the Hayward and Schwarzschild black holes.

In the following, we present the results for $s=1$, corresponding to the PPL mode. The corresponding results for the PPM mode can be obtained in an analogous manner, since the difference between the two polarization modes enters only through the sign of the higher-curvature contributions. Therefore, we expect the PPM mode to exhibit qualitatively similar behavior to that found for the PPL mode.

\section{Time Delay in EFT-Corrected Hayward black hole}
\label{sec:time delay}

In this section, we investigate the time delay of photons in the EFT-corrected Hayward black hole spacetime. As in the gravitational lensing analysis presented in the previous section, we use the time delay as an observable to explore potential differences between the Hayward black hole and the Schwarzschild black hole. In particular, we derive the time delay in the strong-deflection limit and examine how the EFT-induced modifications to photon propagation affect this observable. This analysis allows us to assess whether the time-delay signal can provide an additional observational signature for distinguishing the Hayward black hole from the Schwarzschild black hole.

\subsection{Travel Time Integral}

The coordinate time along a photon trajectory can be obtained by eliminating the affine parameter from the equations of motion,
\begin{equation}
\frac{dt}{dr} = \frac{\dot{t}}{\dot{r}},
\end{equation}
where $\dot{t}$ and $\dot{r}$ are determined by the conserved energy and angular momentum associated with the static and spherically symmetric spacetime. Using these conserved quantities together with the null condition for photon propagation, the radial derivative of the coordinate time can be written as
\begin{align}
\frac{dt}{dr}
=&\dfrac{\sqrt{BA_0}}{\sqrt{A}}
\dfrac{1}{\sqrt{A_0-A\frac{C_0}{C}}}.
\end{align}

The travel time along a photon trajectory is then obtained by integrating this expression from the point of closest approach to the observer and the source. In analogy with the treatment of the deflection angle, it is useful to reorganize the resulting integral so as to separate the contribution associated with the photon-sphere region from the remaining regular part. This decomposition provides a convenient starting point for analyzing the behavior of the travel time in the strong-deflection regime, where the photon trajectory approaches the unstable photon orbit.

The difference in travel times between two photon trajectories characterized by the closest approach radii $r_{0,1}$ and $r_{0,2}$ can be expressed as
\begin{align}
T_1-T_2
=&\tilde{T}(r_{0,1})-\tilde{T}(r_{0,2})
+2\int_{r_{0,1}}^{r_{0,2}}
\frac{\sqrt{B(r)}}{\sqrt{A(r)}}\,dr,
\end{align}
where the functions entering this decomposition are defined by
\begin{align}
\tilde{T}(r_0)
=&\int_0^1
\tilde{R}
\frac{1}{\sqrt{A_0-A(r(z))\frac{C_0}{C(r(z))}}}\,dz,
\\
\tilde{R}(z,r_0)
=&2\frac{r_0}{(1-z)^2}
\frac{\sqrt{B(r(z))A_0}}{\sqrt{A(r(z))}}
\left(
1-
\frac{
\sqrt{A_0-A(r(z))\frac{C_0}{C(r(z))}}
}{
\sqrt{A_0}
}
\right).
\end{align}

Here, the variable $z$ is defined by Eq.~(\ref{zcoor}) and parametrizes the radial integration domain such that the closest approach is mapped to $z=0$. This form is particularly useful because the divergent behavior associated with the photon sphere is contained in the factor involving $A_0-A(r(z))C_0/C(r(z))$.

As the closest approach radius approaches the photon-sphere radius, the leading contribution to the travel-time integral can be extracted by expanding the relevant functions around $z=0$. The resulting expression takes the canonical form
\begin{align}
\tilde{T}(u)
=-\tilde{a}\log\left(\frac{u}{u_{\rm ph}}-1\right)
+\tilde{b}
+\mathcal{O}(u-u_{\rm ph}),
\end{align}
where $u_{\rm ph}$ denotes the critical impact parameter associated with the photon sphere.

The finite coefficient $\tilde{b}$ is obtained after extracting the logarithmically divergent contribution and is given by
\begin{align}
\tilde{b}
=-\pi+\tilde{b}_{R}
+\tilde{a}\log\left(\frac{2q_{\rm ph}}{A_{\rm ph}}\right),
\end{align}
where $\tilde{b}_{R}$ denotes the regular part of the travel-time integral.

The coefficient of the logarithmic term is determined by the behavior of the metric functions in the vicinity of the photon sphere and is given by
\begin{align}
\tilde{a}
=\dfrac{\tilde{R}(0,r_{\rm ph})}{2\sqrt{q_{\rm ph}}}.
\end{align}

Thus, the leading logarithmically divergent contribution to the photon travel time in the strong-deflection limit is determined by the local geometry near the photon sphere, while the constant term $\tilde{b}$ contains the remaining regular contribution.

\subsection{Time Delay in the Hayward Spacetime}

We now apply the above formalism to the four representative values of $g^3$ considered in the previous section. The photon-sphere radius $r_{\rm ph}$, critical impact parameter $u_{\rm ph}$, and strong-deflection coefficients obtained there are used in the following analysis of the photon travel time.

For $g^3=0$, corresponding to the Schwarzschild limit, the strong-deflection time delay can be obtained analytically. For the remaining three
representative values of $g^3$, we numerically evaluate the travel-time integral and focus on its logarithmically divergent contribution in the strong-deflection regime.

The resulting values of the strong-deflection observables for the four cases are summarized in Table~\ref{tab:tilde_a} and \ref{tab:tilde_b}. The $g^3=0$ case is included as the Schwarzschild limit, while the remaining cases illustrate the effects of the nonzero Hayward parameter on the photon-sphere structure and the associated observables.
\begin{align}
\tilde{b}=&-\pi+12-6\sqrt{3}-12\sqrt{3}\arctan(\sqrt{3})+3\sqrt{3}\log6+\sqrt{3}\log46656+\log(58406416-33720960\sqrt{3})\notag\\
&+\frac{8}{3}(-3+\sqrt{3})s\gamma,\\
\tilde{T}(u)=&-3\sqrt{3}\log\left(\frac{u}{3\sqrt{3}-\frac{4s\gamma}{\sqrt{3}}}-1\right)+\tilde{b}+\mathcal{O}(u-u_{\rm ph}).
\end{align}

\begin{figure}[t]
    \centering

    \begin{subfigure}{0.32\textwidth}
        \centering
        \includegraphics[width=\textwidth]{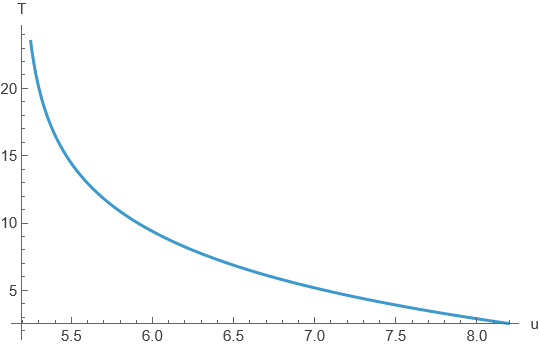}
        \caption{$g^3=0$}
    \end{subfigure}
    \hfill
    \begin{subfigure}{0.32\textwidth}
        \centering
        \includegraphics[width=\textwidth]{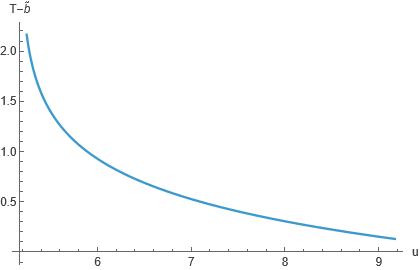}
        \caption{$g^3=\left(\frac{119}{40}\right)^{5/2}\left(\sqrt{3}-2\sqrt{\dfrac{119}{40}}\right)$}
    \end{subfigure}
    \hfill
    \begin{subfigure}{0.32\textwidth}
        \centering
        \includegraphics[width=\textwidth]{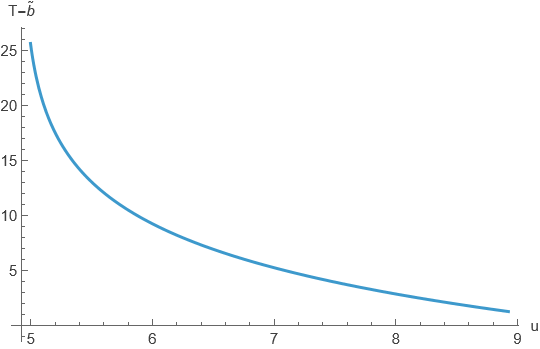}
        \caption{$g^3=\frac{1}{3}\left(\frac{8}{3}\right)^{5/2}(3\sqrt{3}-2\sqrt{6})$}
    \end{subfigure}
    \hfill
    \begin{subfigure}{0.32\textwidth}
        \centering
        \includegraphics[width=\textwidth]{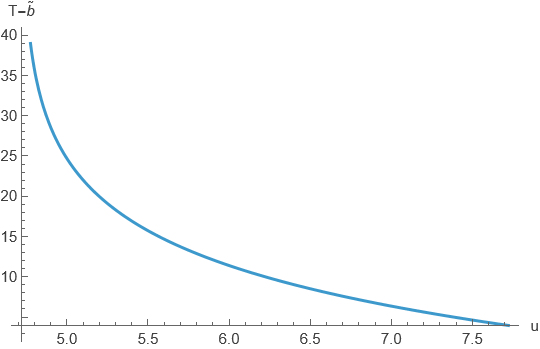}
        \caption{$g^3=\frac{1}{3}\left(\frac{7}{3}\right)^{5/2}(3\sqrt{3}-\sqrt{21})$}
    \end{subfigure}
    \hfill
    \begin{subfigure}{0.32\textwidth}
        \centering
        \includegraphics[width=\textwidth]{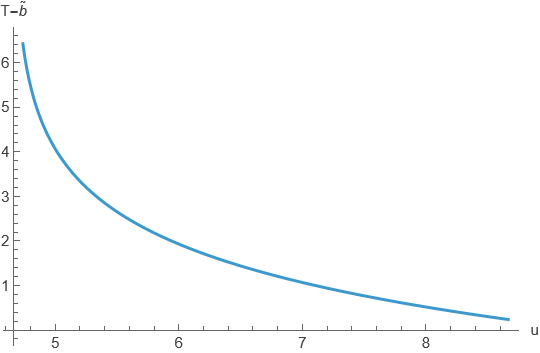}
        \caption{$g^3=\frac{1}{6}\left(\frac{13}{6}\right)^{5/2}(6\sqrt{3}-\sqrt{78})$}
    \end{subfigure}

    \caption{Photon travel time $T(u)$ for the higher-curvature-corrected Hayward black hole with $s=1$ (PPL mode) and higher-curvature coupling constants $\beta=\gamma=10^{-3}$. For $g^3\neq0$, the constant contribution $\tilde{b}$ is subtracted for clarity, whereas for the Schwarzschild case, $g^3=0$, the full time delay is shown without subtracting $\tilde{b}$.}
    \label{fig:RB0_time_delay}
\end{figure}

\begin{figure}[t]
    \centering

     \begin{subfigure}{0.32\textwidth}
        \centering
        \includegraphics[width=\textwidth]{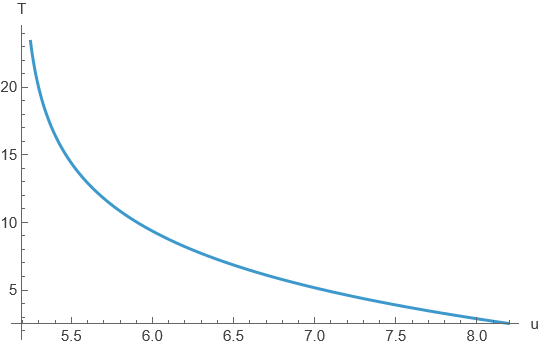}
        \caption{$g^3=0$}
    \end{subfigure}
    \hfill
    \begin{subfigure}{0.32\textwidth}
        \centering
        \includegraphics[width=\textwidth]{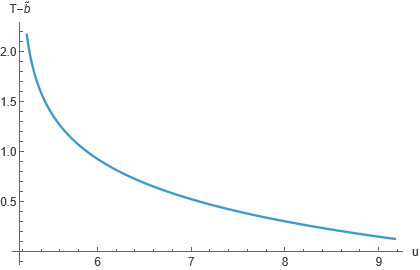}
        \caption{$g^3=\left(\frac{119}{40}\right)^{5/2}\left(\sqrt{3}-2\sqrt{\dfrac{119}{40}}\right)$}
    \end{subfigure}
    \hfill
    \begin{subfigure}{0.32\textwidth}
        \centering
        \includegraphics[width=\textwidth]{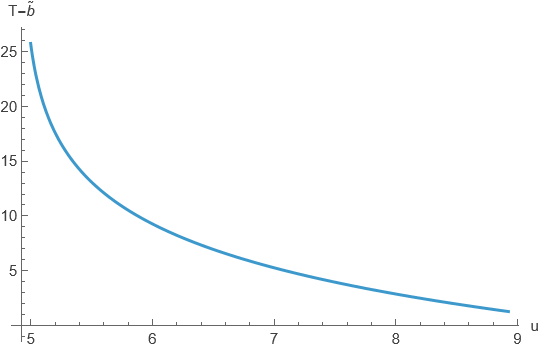}
        \caption{$g^3=\frac{1}{3}\left(\frac{8}{3}\right)^{5/2}(3\sqrt{3}-2\sqrt{6})$}
    \end{subfigure}
    \hfill
    \begin{subfigure}{0.32\textwidth}
        \centering
        \includegraphics[width=\textwidth]{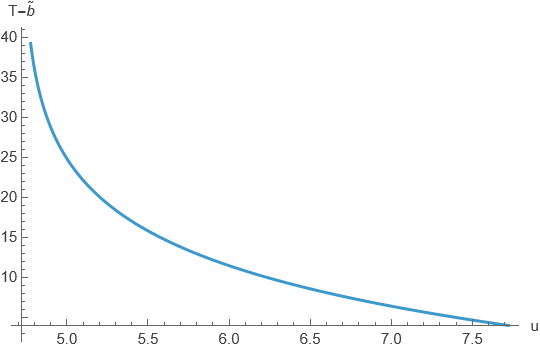}
        \caption{$g^3=\frac{1}{3}\left(\frac{7}{3}\right)^{5/2}(3\sqrt{3}-\sqrt{21})$}
    \end{subfigure}
    \hfill
    \begin{subfigure}{0.32\textwidth}
        \centering
        \includegraphics[width=\textwidth]{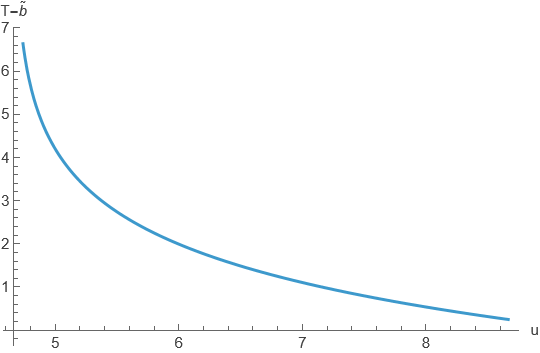}
        \caption{$g^3=\frac{1}{6}\left(\frac{13}{6}\right)^{5/2}(6\sqrt{3}-\sqrt{78})$}
    \end{subfigure}

    \caption{Photon travel time $T(u)$ for the Hayward black hole without EFT corrections. For $g^3\neq0$, the constant term $\tilde{b}$ is subtracted for clarity, whereas for the Schwarzschild case, $g^3=0$, the full time delay is shown without subtracting $\tilde{b}$.}
    \label{fig:RB00_time_delay}
\end{figure}

For each of the five representative values of $g^3$ considered above, we investigate the photon travel time in the strong-deflection regime and compare the results with and without the EFT corrections. For $g^3=0$, corresponding to the Schwarzschild limit, the full strong-deflection expansion of the photon travel time can be obtained analytically. For the nonzero values of $g^3$, we focus on the logarithmically divergent part of the photon travel time, since the contribution of the perturbative EFT corrections to this term may become more pronounced in the strong-deflection regime. The photon travel times including the EFT corrections are shown in Fig.~\ref{fig:RB0_time_delay}, while the corresponding results without the EFT corrections are shown in Fig.~\ref{fig:RB00_time_delay}. The EFT coupling constants are fixed to $\beta=\gamma=10^{-3}$ in the plots.

\begin{figure}[t]
    \centering

    \begin{subfigure}{0.32\textwidth}
        \centering
        \includegraphics[width=\textwidth]{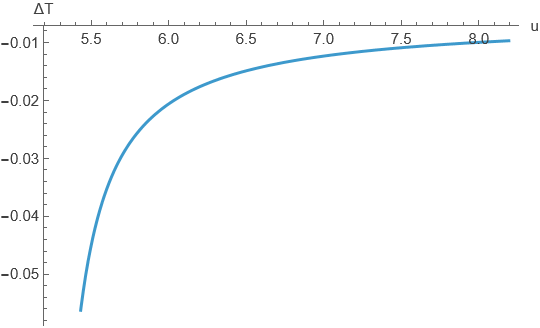}
        \caption{$g^3=0$}
    \end{subfigure}
    \hfill
    \begin{subfigure}{0.32\textwidth}
        \centering
        \includegraphics[width=\textwidth]{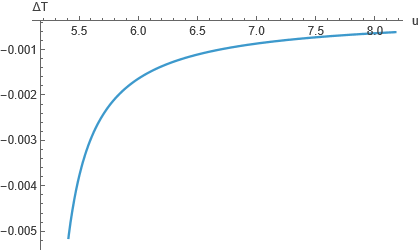}
        \caption{$g^3=\left(\frac{119}{40}\right)^{5/2}\left(\sqrt{3}-2\sqrt{\dfrac{119}{40}}\right)$}
    \end{subfigure}
    \hfill
   \begin{subfigure}{0.32\textwidth}
        \centering
        \includegraphics[width=\textwidth]{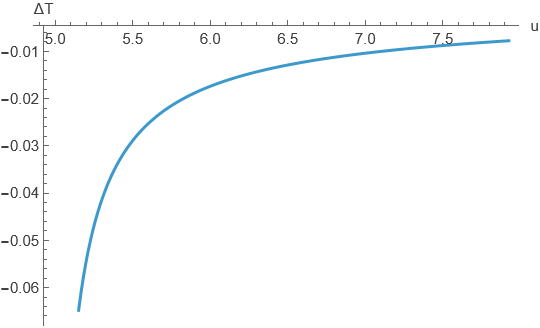}
        \caption{$g^3=\frac{1}{3}\left(\frac{8}{3}\right)^{5/2}(3\sqrt{3}-2\sqrt{6})$}
    \end{subfigure}
    \hfill
    \begin{subfigure}{0.32\textwidth}
        \centering
        \includegraphics[width=\textwidth]{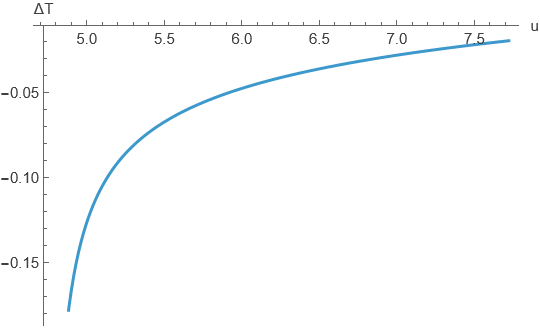}
        \caption{$g^3=\frac{1}{3}\left(\frac{7}{3}\right)^{5/2}(3\sqrt{3}-\sqrt{21})$}
    \end{subfigure}
    \hfill
    \begin{subfigure}{0.32\textwidth}
        \centering
        \includegraphics[width=\textwidth]{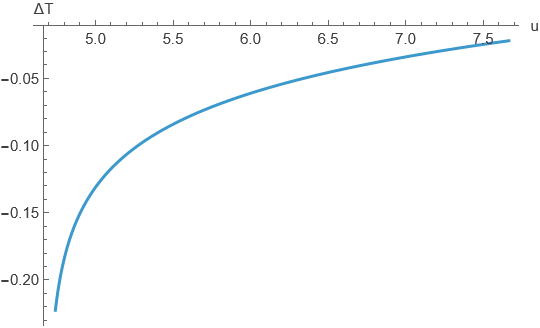}
        \caption{$g^3=\frac{1}{6}\left(\frac{13}{6}\right)^{5/2}(6\sqrt{3}-\sqrt{78})$}
    \end{subfigure}

    \caption{Difference in the photon travel time, $\Delta T(u)$, between the higher-curvature-corrected Hayward black hole and the Hayward black hole without EFT corrections, for the $s=1$ (PPL mode) with higher-curvature coupling constants $\beta=\gamma=10^{-3}$. For $g^3\neq0$, the constant contribution $\tilde{b}$ is subtracted for clarity, whereas for the Schwarzschild case, $g^3=0$, the full time-delay difference is shown without subtracting $\tilde{b}$.}
    \label{fig:RB_time_delay_delta}
\end{figure}

For each value of $g^3$, we also consider the difference between the time delays with and without the EFT corrections,
\begin{align}
\Delta T = T_{\rm EFT}-T_{\rm 0},
\end{align}
where $T_{\rm EFT}$ and $T_{\rm 0}$ denote the photon travel times with and without the EFT corrections, respectively. The difference $\Delta T$ isolates the effect of the EFT corrections at each fixed value of $g^3$. The corresponding results are shown in Fig.~\ref{fig:RB_time_delay_delta}, allowing us to assess the contribution of the EFT corrections to the strong-deflection behavior.

\begin{table}[htbp]
\centering
\caption{Strong-deflection coefficient $\tilde{a}$ for different values of $g^3$, including the leading-order EFT corrections, with $M=1$.}
\label{tab:tilde_a}
\begin{tabular}{c|c}
\hline
$g^3$ & $\tilde{a}$ \\
\hline
$0$ & $3\sqrt{3}$\\
\hline
$\left(\frac{119}{40}\right)^{5/2}\left(\sqrt{3}-2\sqrt{\dfrac{119}{40}}\right)$ & $\sqrt{\dfrac{5}{2(-50+\sqrt{3570})}}-\dfrac{100\sqrt{10}((-354851693+5939084\sqrt{3570})\beta+238(-2164107+36220\sqrt{3570})s\gamma)}{1685159(-50+\sqrt{3570})^{5/2}}$ \\
\hline
 $\frac{1}{3}\left(\frac{8}{3}\right)^{5/2}(3\sqrt{3}-2\sqrt{6})$ & \makecell[l]{
$\dfrac{8}{3}(1+\sqrt{2})\sqrt{\dfrac{1}{7}(2+3\sqrt{2})}-3\sqrt{\dfrac{1}{7}(2+3\sqrt{2})}\dfrac{15(-25408690+17966657\sqrt{2})\beta+98(-7600234+5374177\sqrt{2})s\gamma}{2(-84944954+60065153\sqrt{2})}$} \\
\hline
 $\frac{1}{3}\left(\frac{7}{3}\right)^{5/2}(3\sqrt{3}-\sqrt{21})$ & \makecell[l]{
$\dfrac{7\cdot7^{1/4}}{3\sqrt{24-9\sqrt{7}}}+\dfrac{2}{7^{3/4}(-140+53\sqrt{7})}\sqrt{\frac{8}{3}+\sqrt{7}}\Bigl(3(-2835+1071\sqrt{7}-1090\sqrt{3(7+4\sqrt{7})}$\\
$+412\sqrt{21(7+4\sqrt{7})}+1914\sqrt{21(21+8\sqrt{7})}-5064\sqrt{63+24\sqrt{7}}\Bigr)\beta+2\Bigl(-8295+3135\sqrt{7}$\\
$-2831\sqrt{3(7+4\sqrt{7})}+1070\sqrt{21(7+4\sqrt{7})}+4971\sqrt{21(21+8\sqrt{7})}-13152\sqrt{63+24\sqrt{7}}\Bigr)s\gamma\Bigr)$} \\
\hline
 $\frac{1}{6}\left(\frac{13}{6}\right)^{5/2}(6\sqrt{3}-\sqrt{78})$ & $\dfrac{1}{2\sqrt{-5+\sqrt{26}}}+\dfrac{9((-7293+1373\sqrt{26})\beta+130(-77+15\sqrt{26})s\gamma)}{2197(-5+\sqrt{26})^{3/2}}$ \\
\hline
\end{tabular}
\end{table}

\begin{table}[htbp]
\centering
\caption{Numerical values of the regular contribution $\tilde{b}_R$ and the strong-deflection coefficient $\tilde{b}$ for different values of $g^3$, with the background contribution $\tilde{b}_R^{(0)}$ and the EFT corrections $\tilde{b}_R^{(\beta)}$, $\tilde{b}_R^{(\gamma)}$ proportional to $\beta$ and $s\gamma$ shown separately, with $M=1$.}
\label{tab:tilde_b}
\begin{tabular}{c|c|c|c|c}
\hline
$g^3$ & $\tilde{b}_R^{(0)}$ & $\tilde{b}_R^{(\beta)}$ & $\tilde{b}_R^{(\gamma)}$ & $\tilde{b}$ \\
\hline
$0$ & $-6.48602$ & $0$ & $-8$ & $-0.317358-3.3812s\gamma$\\
\hline
$\left(\frac{119}{40}\right)^{5/2}\left(\sqrt{3}-2\sqrt{\dfrac{119}{40}}\right)$ & -1.2741 & 0.0598 & 0.22074 & $-3.51696+0.05669\beta+0.67145s\gamma$\\
\hline
 $\frac{1}{3}\left(\frac{8}{3}\right)^{5/2}(3\sqrt{3}-2\sqrt{6})$ & -9.0558 & 4.979439 & -6.79394 & $-3.07303+96\beta+100.003s\gamma$\\
\hline
 $\frac{1}{3}\left(\frac{7}{3}\right)^{5/2}(3\sqrt{3}-\sqrt{21})$ & -18.0862 & 78.6884 & 9.36967 & $22.5616+91.8799\beta+21.3379s\gamma$\\
\hline
 $\frac{1}{6}\left(\frac{13}{6}\right)^{5/2}(6\sqrt{3}-\sqrt{78})$ & -5.75204 & 214.343 & 48.3265 & $-9.02816+288.584\beta+66.6942s\gamma$\\
\hline
\end{tabular}
\end{table}

\subsection{Time Delay Between Relativistic Images}

In the strong deflection regime, relativistic images correspond to photon trajectories that wind multiple times around the photon sphere of the wormhole. For trajectories with winding number $n$, the corresponding impact parameters satisfy
\begin{align}
\frac{u}{u_{\rm ph}} - 1 \simeq \exp\left(\frac{\bar{b} - 2\pi n}{\bar{a}}\right).
\end{align}

Substituting this relation into the time expression, we obtain
\begin{align}
T_n = \tilde{a} \frac{2\pi n - \bar{b}}{\bar{a}} + \tilde{b}.
\end{align}
Here, $T_n$ denotes the photon travel time for a trajectory that winds $n$ times around the photon sphere.

The time delay between two relativistic images labeled by $n$ and $m$ is therefore given by
\begin{align}
\Delta T_{n,m}\equiv& \ T_n-T_m\notag\\
=&\ 2\pi(n-m)\left|\frac{\tilde{a}}{\bar{a}}\right|+2\sqrt{\dfrac{B_{\rm ph}}{A_{\rm ph}}}\sqrt{\dfrac{u_{\rm ph}}{c}}e^{\frac{\bar{b}}{2\bar{a}}}\left(e^{-\frac{\pi n}{|\bar{a}|}}-e^{\frac{\pi m}{|\bar{a}|}}\right)\notag\\
\equiv&\ \tilde{T}_{1}(n-m)+\tilde{T}_{2}\left(e^{-\frac{\pi n}{|\bar{a}|}}-e^{\frac{\pi m}{|\bar{a}|}}\right),
\end{align}
where $\tilde{T}_{1}$ and $\tilde{T}_{2}$ are defined by
\begin{align}
\tilde{T}_{1}=&2\pi\left|\frac{\tilde{a}}{\bar{a}}\right|,\\
\tilde{T}_{2}=&2\sqrt{\dfrac{B_{\rm ph}}{A_{\rm ph}}}\sqrt{\dfrac{u_{\rm ph}}{c}}e^{\frac{\bar{b}}{2\bar{a}}},
\end{align}
where $c$ is the quantity efined in Eq.~(\ref{c_def}).

Below, we present the numerical results for the time delay in the EFT-corrected Hayward black hole spacetime. For simplicity, we set the mass parameter to $M=1$. The quantities $\tilde{T}_1$ and $\tilde{T}_2$ for both the PPL and PPM modes are obtained numerically for representative values of the Hayward parameter $g^3$ and are summarized in Tables~\ref{tab:T1} and~\ref{tab:T2}.

These results show that, for both the PPL and PPM modes, the time delay retains the same structural dependence on the winding numbers $(n,m)$ and the strong-deflection coefficients. The polarization dependence enters through the EFT correction terms, with the PPL and PPM modes differing in the sign of the corresponding Weyl-curvature contribution. However, the time delay itself is not particularly sensitive to the perturbatively small EFT corrections. Therefore, it may be difficult to distinguish the Hayward black hole from the Schwarzschild black hole solely through the EFT-induced corrections to the time-delay observable.

The photon travel time, on the other hand, exhibits an important feature in the strong-deflection regime. For photon trajectories with the impact parameter approaching its critical value, the photon travel time develops a logarithmic divergence, analogous to that of the deflection angle. This divergence is associated with the increasing number of windings around the unstable photon orbit and reflects the prolonged propagation of photons near the photon sphere. Although the photon travel time itself is not a directly observable quantity, this behavior characterizes the winding of photons near the critical orbit. The observable time delay is obtained from the difference between the travel times of photons following different trajectories and therefore retains information about their winding numbers. Importantly, although the EFT corrections considered here are perturbatively small, their contributions to the photon travel time accumulate with the winding number and can become enhanced for sufficiently large winding numbers. Thus, the time-delay observable may retain signatures of these enhanced EFT effects and provide additional information for distinguishing the Hayward black hole from the Schwarzschild black hole. A combined analysis of the deflection angle and time delay may therefore provide a more complete characterization of the differences between the two spacetimes.

\begin{table}[htbp]
\centering
\caption{Values of $\tilde{T}_1$ for representative values of the parameter $g^3$.}
\label{tab:T1}
\begin{tabular}{c|c}
\hline
$g^3$ & $\tilde{T}_1$ \\
\hline
$0$ & $6\sqrt{3}\pi-\dfrac{8s\gamma\pi}{\sqrt{3}}$ \\
\hline
$\left(\frac{119}{40}\right)^{5/2}\left(\sqrt{3}-2\sqrt{\dfrac{119}{40}}\right)$ & $\pi-\dfrac{3200((-10018610+167673\sqrt{3570})\beta+119(-59980+1007\sqrt{3570})s\gamma)\pi}{1685159(-50+\sqrt{3570})^2}$ \\
\hline
$\dfrac{1}{3}\left(\dfrac{8}{3}\right)(3\sqrt{3}-2\sqrt{6})$ & \makecell[l]{
$\frac{16}{3}(1+\sqrt{2})\sqrt{2-\sqrt{2}}\pi$\\
$-\dfrac{3(\sqrt{2-\sqrt{2}}((-2185457156+1545351575\sqrt{2})\beta+4(-905245994+640105581\sqrt{2})s\gamma) \pi}{1810491988-1280211162\sqrt{2}}$} \\
\hline
$\dfrac{1}{3}\left(\dfrac{7}{3}\right)(3\sqrt{3}-\sqrt{21})$ &  \makecell[l]{
$\frac{14}{3}\dfrac{7^{1/4}\pi}{\sqrt{-2+\sqrt{7}}}-\dfrac{4\pi}{7\sqrt{-37+14\sqrt{7}}(-140+53\sqrt{7})}\Bigl(3\cdot 7^{1/4}\sqrt{8-3\sqrt{7}}\Bigl(-791+299\sqrt{7})\beta+6\sqrt{3}(6699-2532\sqrt{7}$\\
$+7^{1/4}\sqrt{-28+11\sqrt{7}}(-545+206\sqrt{7}\Bigr)\beta+18\cdot7^{1/4}\sqrt{8-3\sqrt{7}}\left(-217+82\sqrt{7}\right)s\gamma+2\sqrt{3}\Bigl(34797-13152\sqrt{7}$  \\
$+7^{1/4}\sqrt{-28+11\sqrt{7}}(-2831+1070\sqrt{7})s\gamma\Bigr)$}\\
\hline
$\dfrac{1}{6}\left(\dfrac{13}{6}\right)(6\sqrt{3}-\sqrt{78})$ & $\pi+\dfrac{((-91+16\sqrt{26})\beta+13(-4+\sqrt{26})s\gamma)\pi}{2197}$\\
\hline
\end{tabular}
\end{table}
\begin{table}[htbp]
\centering
\caption{Values of $\tilde{T}_2$ for representative values of the parameter $g^3$.}
\label{tab:T2}
\begin{tabular}{c|c}
\hline
$g^3$ & $\tilde{T}_2$ \\
\hline
$0$ & \makecell[l]{
$6\sqrt{6}e^{\frac{1}{2}\left(-\pi-4\arctan\left(\frac{1}{\sqrt{3}}\right)+\log216\right)}(1+\frac{1}{45}s\gamma)$\\
$\times\left(-26+8\sqrt{3}+5\pi+20\arctan\left(\frac{1}{\sqrt{3}}\right)+\log1024-20\log(1+\sqrt{3})\right)$}\\
\hline
$\left(\frac{119}{40}\right)^{5/2}\left(\sqrt{3}-2\sqrt{\dfrac{119}{40}}\right)$ & $12.0851+1.38747\beta+1.90453s\gamma$ \\
\hline
$\dfrac{1}{3}\left(\dfrac{8}{3}\right)(3\sqrt{3}-2\sqrt{6})$ & $12.6754-0.936583\beta+1.89974s\gamma$ \\
\hline
$\dfrac{1}{3}\left(\dfrac{7}{3}\right)(3\sqrt{3}-\sqrt{21})$ & $11.9397+1542.14\beta$ \\
\hline
$\dfrac{1}{6}\left(\dfrac{13}{6}\right)(6\sqrt{3}-\sqrt{78})$ & $7.90589+200.738\beta-176.221s\gamma$
\\
\hline
\end{tabular}
\end{table}

\section{Conclusion and Discussion}
\label{sec:conclusion}

In this work, we have investigated whether a Hayward regular black hole can be observationally distinguished from a Schwarzschild black hole through strong gravitational lensing when effective field theory (EFT) corrections to photon propagation are taken into account. Starting from an effective action containing curvature-photon couplings, we derived the modified propagation law for photons and constructed the corresponding effective metric for the polarization-dependent photon trajectories. This provides a framework for studying gravitational lensing in regular black hole spacetimes while consistently incorporating EFT corrections to photon propagation.

We first analyzed the photon trajectories in the strong-deflection regime and derived the corresponding photon-sphere radius, critical impact parameter, and strong-deflection coefficients. Although the EFT corrections are parametrically small, their effects can become enhanced in the strong-deflection limit, where the deflection angle exhibits a logarithmic divergence as the impact parameter approaches its critical value. We found that the EFT corrections modify the strong-deflection observables and therefore leave characteristic imprints on the gravitational lensing of photons. By evaluating these observables for representative values of the Hayward parameter and comparing them with the Schwarzschild case, we showed that the combined effects of the regular black hole geometry and the modified photon propagation can lead to distinguishable features in the strong-lensing observables.

An important feature of the present analysis is that the EFT corrections do not simply produce a universal shift of the lensing observables. Rather, their contributions depend on the curvature structure of the underlying spacetime. Since the Hayward black hole is not Ricci-flat, curvature-photon interactions that vanish in the Schwarzschild spacetime can contribute to photon propagation in the Hayward geometry. Consequently, the EFT corrections can encode information about the non-vacuum curvature structure of the regular black hole and may enhance the difference between the Hayward and Schwarzschild predictions. This demonstrates that strong gravitational lensing can provide a sensitive probe of both the background geometry and curvature-dependent modifications to photon propagation.

We have also investigated the photon time delay as a complementary observable. In the strong-deflection regime, the photon travel time itself exhibits a logarithmic divergence as the impact parameter approaches its critical value. This logarithmically divergent contribution corresponds to the increasingly large winding number of photons around the unstable photon orbit and reflects the prolonged propagation of photons in the vicinity of the photon sphere. Although the photon travel time itself is not a directly observable quantity, the time delay is constructed from differences between the travel times of photons with different winding numbers. The logarithmically divergent contribution to the photon travel time therefore provides the underlying origin of the winding-dependent time-delay signal. Importantly, although the EFT corrections considered here are perturbatively small, their contributions to the photon travel time accumulate with the winding number and can therefore become enhanced for sufficiently large winding numbers. This suggests that the corresponding EFT effects may be reflected in the time-delay observable and may provide additional information for distinguishing the Hayward black hole from the Schwarzschild black hole. A combined analysis of the deflection angle and time delay may therefore provide a more complete characterization of the differences between the two spacetimes and their potential observational distinguishability.

The results obtained here suggest that strong gravitational lensing can provide a promising observational probe for distinguishing regular black holes from Schwarzschild black holes, even when the corrections to photon propagation are perturbatively small. In particular, the logarithmic enhancement of the deflection angle near the critical propagation region provides a mechanism through which small EFT effects can leave potentially observable signatures. We also find that the contribution of the EFT corrections becomes more pronounced as the Hayward parameter $g^3$ approaches its critical value, indicating that the effects of curvature-dependent corrections to photon propagation can be particularly significant in the near-critical regime. This behavior is especially relevant for testing regular black hole models, for which deviations from the Schwarzschild geometry may otherwise be difficult to identify.

Several directions remain for future investigation. First, it would be interesting to extend the present analysis to other regular black hole models, such as the Bardeen \cite{1968qtr..conf...87B} and Ay\'on-Beato--Garc\'ia black holes \cite{Ayon-Beato:1998hmi,Ayon-Beato:1999kuh}, and investigate how curvature-dependent EFT corrections to photon propagation manifest themselves in their strong-lensing observables. Such an extension would allow us to systematically examine how the strong-deflection observables depend on the specific structure of different regular black hole geometries and to identify characteristic features that distinguish them from the Schwarzschild case. Second, a more realistic treatment should incorporate finite-distance effects, higher-order terms in the strong-deflection expansion, and a systematic exploration of the allowed parameter space of the Hayward and EFT parameters. It would also be interesting to investigate whether the characteristic signatures identified here can be constrained by current or future high-precision gravitational-lensing observations. These extensions may clarify the potential of strong gravitational lensing as a probe of regular black hole geometries and curvature-dependent corrections to photon propagation.

\bibliography{references}

\end{document}